\documentclass[usenatbib]{mnras}

\usepackage{lscape}
\usepackage{graphicx}
\usepackage{amssymb}
\usepackage{amsmath}
\usepackage{url}
\usepackage{bm}
\usepackage{aas_macros}
\usepackage{rotating}

\usepackage{multirow}

\title[Intrinsic reddening of the Magellanic Clouds]{The intrinsic reddening of the Magellanic Clouds as traced by background galaxies -- III.
The Large Magellanic Cloud}

\author[C.~P.~M.~Bell et al.]{Cameron~P.~M.~Bell,${^1}$\thanks{E-mail:
  cbell@aip.de (CPMB)} Maria-Rosa L. Cioni,$^{1}$ Angus H.~Wright,$^{2}$ David~L.~Nidever,$^{3,4}$ I-Da Chiang,$^{5,6}$
  \newauthor Samyaday Choudhury,$^{7,8}$ Martin A. T. Groenewegen,$^{9}$ Clara~M.~Pennock,$^{10}$ Yumi Choi,$^{11}$ 
  \newauthor Richard de Grijs,$^{7,8}$ Valentin D. Ivanov,$^{12}$ 
  Pol Massana,$^{13}$ Ambra Nanni,$^{14}$, Noelia E. D. No{\"e}l,$^{13}$ 
  \newauthor Knut Olsen,$^{4}$ Jacco Th. van Loon,$^{10}$ A. Katherina Vivas$^{15}$ and Dennis Zaritsky$^{16}$
  \\
  $^{1}$Leibniz-Institut f{\"u}r Astrophysik Potsdam (AIP), An der Sternwarte 16, D-14482 Potsdam, Germany\\
  $^{2}$Astronomisches Institut, Ruhr-Universit{\"a}t Bochum, Universit{\"a}tsstr. 150, D-44801 Bochum, Germany\\
  $^{3}$Department of Physics, Montana State University, P.O. Box 173840, Bozeman, MT 59717, USA\\
  $^{4}$National Optical--Infrared Astronomy Research Laboratory (NOIRLab), 950 North Cherry Avenue, Tucson, AZ 85719, USA\\
  $^{5}$Institute of Astronomy and Astrophysics, Academia Sinica, No. 1, Section 4, Roosevelt Road, Taipei 10617, Taiwan\\
  $^{6}$Center for Astrophysics and Space Sciences, Department of Physics, University of California, San Diego 9500 Gilman Drive, La Jolla,\\
  CA 92093, USA\\
  $^{7}$Department of Physics and Astronomy, Macquarie University, Balaclava Road, Sydney NSW 2109, Australia\\
  $^{8}$Research Centre for Astronomy, Astrophysics and Astrophotonics, Macquarie University, Balaclava Road, Sydney NSW 2109, Australia\\
  $^{9}$Koninklijke Sterrewacht van Belgi\"{e}, Ringlaan 3, 1180 Brussels, Belgium\\
  $^{10}$Lennard-Jones Laboratories, School of Chemical and Physical Sciences, Keele University, Keele, ST5 5BG, UK\\
  $^{11}$Space Telescope Science Institute, 3700 San Martin Drive, Baltimore, MD 21218, USA\\
  $^{12}$European Southern Observatory, Karl-Schwarzschild-Str. 2, D-85748 Garching bei M{\"u}nchen, Germany\\
  $^{13}$Department of Physics, University of Surrey, Guildford, GU2 7XH, UK\\
  $^{14}$National Centre for Nuclear Research, ul. Pasteura 7, 02-093 Warszawa, Poland\\
  $^{15}$Cerro Tololo Inter-American Observatory, National Optical Astronomy Observatory, Casilla 603, La Serena, Chile\\
  $^{16}$Steward Observatory and Department of Astronomy, The University of Arizona, Tucson, AZ85721, USA\\
}
\begin{document}

\date{Accepted 2022 May 27. Received 2022 May 27; in original form 2022 January 10 }

\pagerange{\pageref{firstpage}--\pageref{lastpage}} \pubyear{2022}

\maketitle

\label{firstpage}

\begin{abstract}
We present a map of the total intrinsic reddening across $\simeq$\,90\,deg$^{2}$ of the Large Magellanic Cloud (LMC) derived
using optical ($ugriz$) and near-infrared (IR; $YJK_{\rm{s}}$) spectral energy distributions (SEDs) of background galaxies. The
reddening map is created from a sample of 222,752 early-type galaxies based on
the \textsc{lephare} $\chi^{2}$ minimisation SED-fitting routine. We find excellent agreement between the regions of enhanced
intrinsic reddening across the central ($4\times4$ deg$^2$) region of the LMC and the morphology of the low-level pervasive dust emission as traced by far-IR emission. In addition, we are able to distinguish smaller, isolated enhancements that are coincident with known star-forming regions and the clustering of young stars observed in morphology maps. The level of reddening associated with the molecular ridge south of 30 Doradus is, however, smaller than in the literature reddening maps. The reduced number of galaxies detected in this region, due to high extinction and crowding, may bias our results towards lower reddening values. Our map is consistent with maps derived from red clump stars and from the analysis of the star formation history across the LMC.
This study represents one of the first large-scale categorisations of extragalactic sources behind the LMC and as such we provide
the \textsc{lephare} outputs for our full sample of $\sim$\,2.5 million sources.
\end{abstract}


\section{Introduction}
\label{introduction}

The Magellanic Clouds contain stellar populations spanning all ages which thanks to their proximity ($\sim 50-60$\,kpc) have allowed a number of investigations. The study of star formation at lower ($0.2-0.5$\,Z$_{\odot}$) metallicities (e.g. \citealp*{Gouliermis14}; \citealp{Zivkov18}), the extragalactic distance scale (and by extension the Hubble constant $H_{0}$, e.g. \citealp*{deGrijs14}; \citealp{deGrijs15}; \citealp{Riess19,Freedman20}), and the metallicity dependence of the period--luminosity relation of variable stars (see e.g. \citealp{Gieren18,Groenewegen18,Muraveva18,Ripepi19}) are examples of  investigations dependent upon an understanding of the amount and spatial distribution of dust across the galaxies. These studies use stellar populations of different ages to quantify the reddening in the Magellanic Clouds which results in statistically significant differences (this is further complicated by variations in metallicity, three-dimensional distributions of the stars and dust, etc.). On average, reddening maps based on young stellar populations such as Classical Cepheids or young open clusters produce maps emphasising the high levels of reddening typical of those areas of ongoing or recent star formation (\citealp{Nayak18,Joshi19}); on the contrary, maps based on older stellar populations such as red clump and RR Lyrae stars systematically result in lower levels of reddening (see e.g. \citealp{Zaritsky02,Muraveva18,Gorski20,Cusano21}). 

The differences in colour excess, $\Delta E(B-V)$, can have a large influence on the inferred value for the Hubble constant and on the three-dimensional structure derived from stars (e.g. \citealp{Ripepi17,Choi18}). For example, an $\Delta E(B-V)=0.06$\,mag difference between \cite*{Haschke11} and \cite{Gorski20} studies, both using data from the third phase of the Optical Gravitational Lensing Experiment (OGLE-III; \citealp{Udalski03}), implies a difference in the value of the Hubble constant larger then 3\,km\,s$^{-1}$\,Mpc$^{-1}$ (see e.g. section~3.1 of \citealp{Riess09}). The presence of dust influences the location of stars in colour--magnitude diagram (CMD) which is often used to select different stellar types to study their spatial distribution (e.g. \citealp{ElYoussoufi19}) and kinematics (e.g. Niederhofer et al., submitted). An independent map of the internal reddening of the Magellanic Clouds would also reduce the residuals between data and models in regions of the CMD which are used simultaneously to infer the reddening, distance modulus and star formation history (SFH) of the galaxies (e.g. \citealp{Rubele18,Mazzi21}). Small-scale reddening variations present in maps obtained from individual stars (e.g. \citealp{Tatton21}) and in specific regions of the galaxies (e.g. \citealp*{deMarchi21}) may not be easy to reconcile with large-scale studies based on broadly distributed tracers.

In \citet[hereafter Paper~I]{Bell19} we piloted a method to map the intrinsic reddening of a foreground extinguishing medium using the spectral energy distributions (SEDs) of background galaxies. This approach takes away the dependency on reddening values of stellar populations and allows us to study the total reddening by sampling the full column depth of the extinguishing medium. In \citet[hereafter Paper~II]{Bell20} we applied this method to a $\simeq$\,34\,deg$^{2}$ region of the Small Magellanic Cloud (SMC) covered by both the Survey of the MAgellanic Stellar History (SMASH; \citealp{Nidever17}) and the Visible and Infrared Survey Telescope for Astronomy (VISTA) survey of the Magellanic Clouds system (VMC; \citealp{Cioni11}). In Paper II, adopting galaxies with low intrinsic reddening, we found signs of higher intrinsic reddening in the main body of the galaxy compared with its outskirts. Our maps agree with those in the literature which were created using young stars whereas there are some discrepancies with maps derived from longer-wavelength far-IR emission and from old stars. These discrepancies may be due to biases in the sample or uncertainties in the far-IR emissivity of the optical properties of the dust grains. In this study, we map the total intrinsic reddening across a  $\simeq$\,90\,deg$^{2}$ region of the Large Magellanic Cloud (LMC) combining observations from the SMASH and VMC surveys.

The structure of this paper is as follows. In Section~\ref{creating_fitting_seds_of_galaxies} we describe the processes involved in
creating and fitting the SEDs of background galaxies as well as provide a comparison of the results of the SED fitting with other
extragalactic studies. The details of how we create our intrinsic reddening map of the LMC are presented in
Section~\ref{intrinsic_reddening_lmc}. We discuss our reddening map and compare it with literature reddening maps of the LMC based on different tracers in Section~\ref{discussion}, and summarise our findings in Section~\ref{summary}.

\section{Creating and fitting SEDs of galaxies}
\label{creating_fitting_seds_of_galaxies}

\begin{figure}
\centering
\includegraphics[width=\columnwidth]{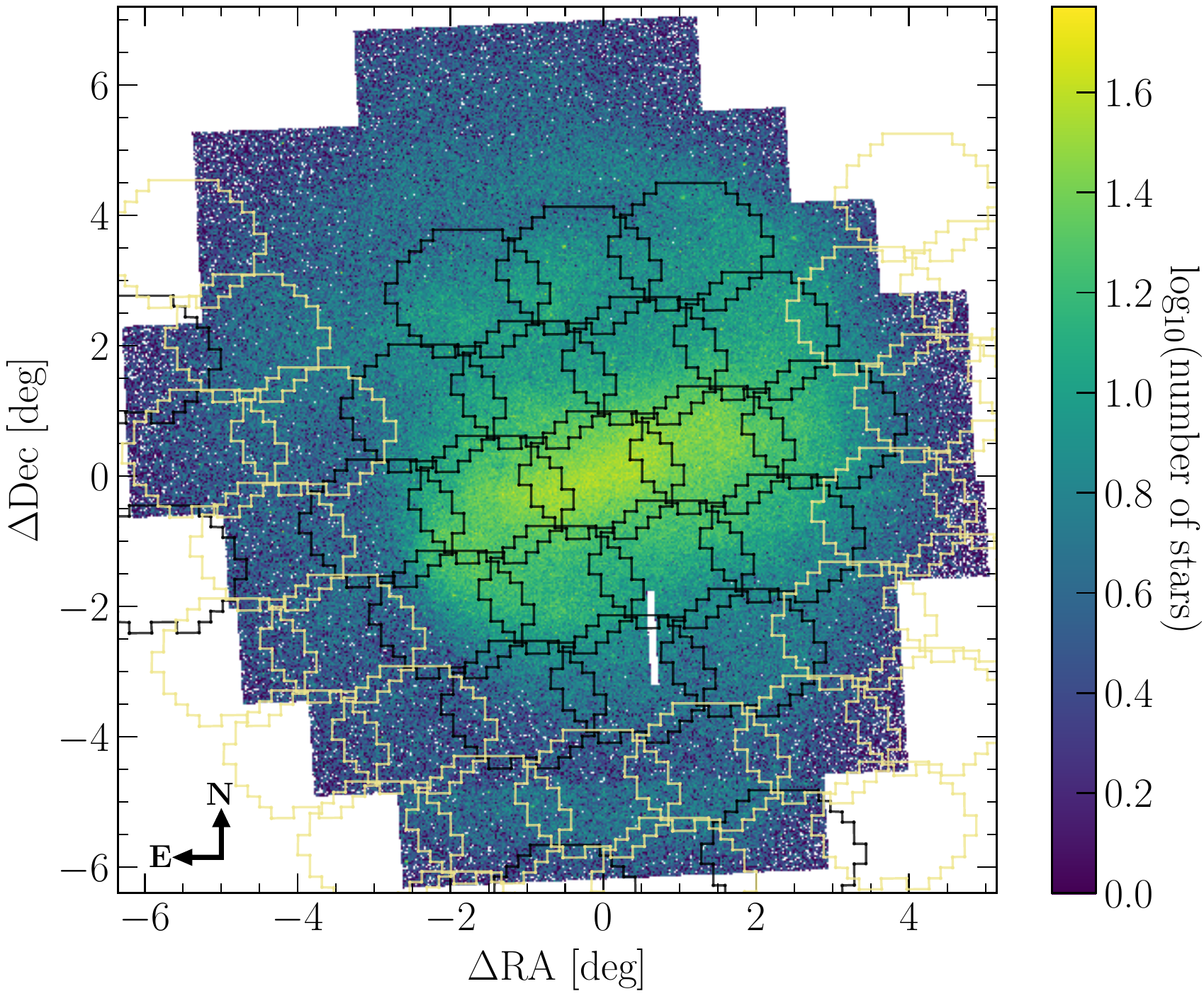}
\caption[]{Stellar density distribution of the LMC from the VMC survey. The colour bar refers to the number of stars per bin, where the bin size is of $0.4\times0.4$ arcmin$^2$. The origin of the plot corresponds to the centre of the LMC
as given by the HYPERLEDA catalogue of \cite{Paturel03}. The hexagons denote the positions of the SMASH fields that overlap with
the VMC coverage of the LMC. Black fields denote those for which $ugriz$ data are available, whereas khaki fields represent those for
which only $griz$ data are available. The combined SMASH--VMC footprint is $\simeq$\,90\,deg$^{2}$. A gap at 
$\Delta$RA\,$\simeq0.7$\,deg is currently being observed.}
\label{fig:smash_vmc_coverage}
\end{figure}

Our data set consists of optical $ugriz$ and near-IR $YJK_{\rm{s}}$ photometry taken as part of the SMASH and VMC surveys,
respectively, that cover the wavelength range 0.3--2.5\,$\mu$m. The combined SMASH--VMC footprint of the LMC is shown in
Fig.~\ref{fig:smash_vmc_coverage} and covers an area of $\simeq$\,90\,deg$^{2}$. We note two significant differences regarding
the SMASH data used in this study compared to that used in Papers~I and II. First, as shown by the different colour DECam fields in
Fig.~\ref{fig:smash_vmc_coverage}, only the central regions of the LMC (black hexagons) are fully covered by deep $ugriz$
data. The surrounding fields (khaki hexagons) represent fields for which only $griz$ data are available. Second, whereas the
fields with full 5-band data consist of deep exposures (999\,s in $uiz$ and 801\,s in $gr$), in addition to three sets of short
60\,s exposures in $ugriz$ with large half-CCD offsets between each set (to cover the gaps between the CCDs), fields with 4-band
data consist only of two sets of short 60\,s exposures in $griz$ (again with half-CCD offsets to cover the gaps between CCDs; see
section~3 of \citealp{Nidever17} for details).

For the study of the LMC we have made three notable changes, relative to the methodology applied to the SMC and laid out in Papers~I and II; namely the selection of background galaxies, the treatment of foreground Milky Way (MW) reddening and the use of ancillary radio data to separate the Active Galactic Nuclei (AGN) population prior to fitting the SEDs. We refer the reader to Paper I of this series for a discussion of each step of the process, including the choice of galaxy templates, redshift priors, extinction law, and photometric uncertainties.

\subsection{Selection of background galaxies}
\label{selection_of_background_galaxies}

\begin{figure*}
\centering
\includegraphics[width=\textwidth]{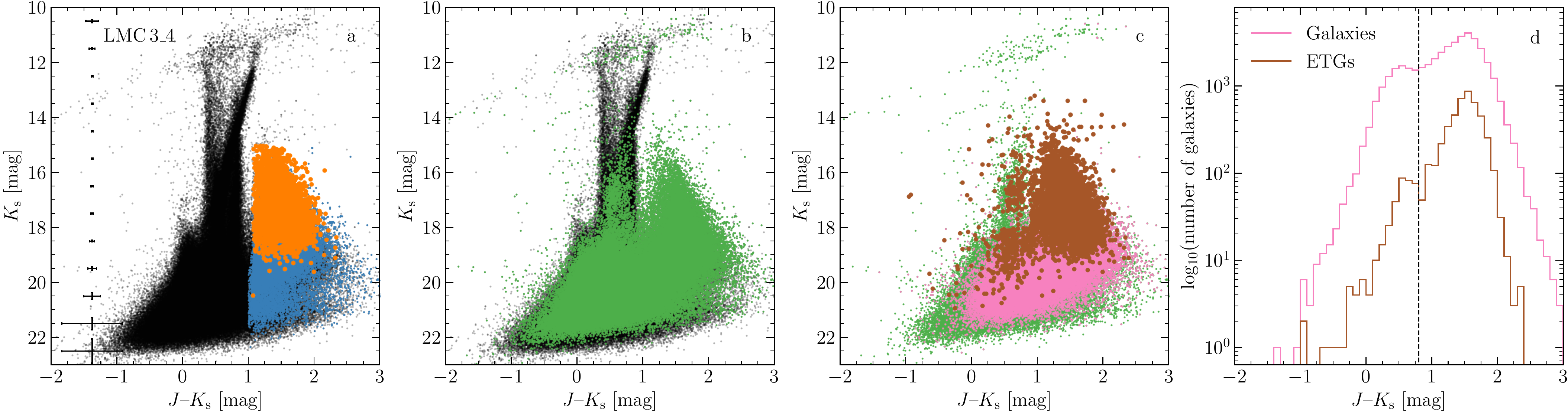} \caption[]{\textbf{a}: Near-IR $J-K_{\rm{s}}, K_{\rm{s}}$ CMD
of tile LMC 3\_4. The blue points denote objects selected as likely galaxies adopting the criteria of Papers~I and II (see
text), whereas the orange points represent those that are  subsequently classified by \textsc{lephare} as ETGs. The error bars
on the left denote the median photometric errors as a function of $K_{\mathrm{s}}$-band magnitude at 1\,mag intervals
\textbf{b}: As \textbf{a}, but the green points correspond to those for which the $K_{\rm{s}}$-band sharpness index is
greater than 0.5. \textbf{c}: Same as \textbf{b}, however the pink points denote those of the green points \textsc{lephare}
classifies as galaxies and the brown points represent the subset of these classified as ETGs. \textbf{d}: Histograms of
the numbers of galaxies and ETGs as a function of $J-K_{\rm{s}}$ colour. The dashed line at a colour of $J-K_{\rm{s}}=0.8$\,mag
separates those that appear to be primarily blended stellar sources (at bluer colours) and those that are predominantly non-blended, extended
sources (at redder colours).}
\label{fig:galaxy_selection}
\end{figure*}

The process of creating the galaxy SEDs has been extensively covered in Papers~I and II. However, in light of our SMC analysis we have made some minor modifications that we describe below. In Paper~II we noted that the number of early-type galaxies (ETGs) in the central regions of the SMC was limited due to incompleteness arising from the enhanced levels of crowding compared to the less-dense outer regions. The central regions of the LMC are even more affected by crowding and thus we have revised our selection in an effort to increase the number of ETGs in our sample.

Before explaining our updated selection criteria, it should be noted that our definition of an ETG is directly linked to the adopted templates in \textsc{lephare}. The galaxy templates we adopt are the so-called AVEROI\_NEW templates and we classify an object as an ETG if \textsc{lephare} finds a lower minimum $\chi^{2}$ associated with the best-fitting galaxy template (compared to the best-fitting stellar template) \emph{and} if the best-fitting galaxy template corresponds to a spectral type of Ell or Sbc (see section~3.1 of Paper~I for details regarding the AVEROI\_NEW templates).

Our selection of background galaxies is based on the VMC point-spread function (PSF) photometric catalogues (see \citealp{Mazzi21} for details). Panel a of Fig.~\ref{fig:galaxy_selection} shows the CMD selection previously adopted in Papers~I and II, namely retaining those that satisfy the following: $J-K_{\rm{s}} > 1$\,mag, $K_{\rm{s}} > 15$\,mag, an associated
stellar probability of less than 0.33 and a $K_{\rm{s}}$-band sharpness index greater than 0.5. To increase our sample of ETGs we effectively need to include galaxies at bluer colours. If we simply remove the colour cut, however, the use of the stellar probability still severely limits the inclusion of bluer galaxies (see e.g. Fig.~2 of Paper~I).

To assess how best to modify our galaxy selection, we first retain only those objects with a $K_{\rm{s}}$-band sharpness index greater than 0.5 (Fig.~\ref{fig:galaxy_selection}b). These objects are then processed through \textsc{lambdar} (see Section~\ref{lambdar_photometry}) and \textsc{lephare} (see Section~\ref{fittings_seds_lephare}) to determine which  ones are classified as galaxies and, more specifically for our purposes, ETGs (the brown and purple points in Fig.~\ref{fig:galaxy_selection}c, respectively). There is clearly a bimodality in the distribution of galaxies and ETGs as a function of $J-K_{\rm{s}}$ colour. Dividing the ETG sample at a colour of $J-K_{\rm{s}}=0.8$\,mag and visually inspecting $K_{\rm{s}}$-band image cut-outs we note that the vast majority of objects at bluer colours (of which there are 369 in tile LMC 3\_4, a tile in the outer disc of the LMC) are stellar blends, whereas those at redder colours (of which there are 4430) appear to be predominantly non-blended extended sources. Retaining such a large number of stellar blends in our ETG sample would have a non-negligible effect when creating the reddening maps. Furthermore, it is not feasible to visually inspect all ($\sim$300,000) candidate ETGs across the combined SMASH--VMC coverage of the LMC. The $J-K_{\rm{s}}$ colour distribution of known QSOs, which is similar to that of galaxies, shows a clear edge at $0.8$ mag which is also away from stellar sources \citep{Cioni13}.
The final number of objects classified as ETGs in tile LMC 3\_4 based on a revised colour selection criteria ($J-K_{\rm{s}} > 0.8$\,mag) is 4430, which when compared to the 3361 one would find using our previous selection criteria represents an increase in sample size of almost a third.

\subsection{{\sc{lambdar}} photometry}
\label{lambdar_photometry}

Fluxes for each of our targets are measured using the Lambda Adaptive Multi-Band Deblending Algorithm in R
(\textsc{lambdar}\footnote{\url{https://github.com/AngusWright/LAMBDAR}}, v0.20.7;
\citealp{Wright16}). This version has minor developments (with respect to v0.20.5 that was used in Paper~II), that improve the
stability and fidelity of the PSF estimation procedure. We use these new developments (and our knowledge that our fields are
crowded with point sources) to specify conservative requirements on which sources are selected for use in the PSF estimation.
Specifically, we opt for a maximum number of point sources to be used in the PSF estimation of 500 per-chip (a chip corresponds to a detector or a charge coupled device in the VISTA and SMASH images, respectively), with a minimal radial
separation (from possible contaminating sources) of 10 pixels (or $\sim$\,$3.4$\,arcsec). The latter condition often produces many
more than 500 sources per-chip (especially in very crowded regions), and in these cases the code iteratively increases the
minimum radial separation until 500 sources are selected; that is, the code selects the 500 most isolated point sources for
use in the PSF estimation.

Paper~I provides a comprehensive discussion regarding the measurement of fluxes in addition to the calibration of both the optical
and near-IR data sets onto an AB magnitude system. We adopt the fluxes measured using a default circular aperture of diameter 3\,arcsec.
The $Y$-, $J$- and $K_{\rm{s}}$-band deep stacks from which \textsc{lambdar} measures the fluxes were downloaded from the VISTA Science
Archive (VSA; \citealp{Cross12}) and have been processed by the Cambridge Astronomy Survey Unit (CASU). Unlike the deep stacks used in
Papers~I and II that were created using the VISTA Data Flow System (VDFS; \citealp{Irwin04}) v1.3, the deep stacks used in this study
have been processed using v1.5 (\citealp{Gonzalez-Fernandez18}; see also the CASU 
webpage\footnote{\url{http://casu.ast.cam.ac.uk/surveys-projects/vista/data-processing/version-log}}). The main difference between the
two versions is in the determination of the zero-points and as such we apply the offsets listed in eqs.~C7--C11 of
\cite{Gonzalez-Fernandez18} in addition to those noted in section~2.4 of Paper~I. The zero-point differences between the two versions amount to $0.01-0.02$ mag which will have a minimal influence on our study (cf. Fig. \ref{fig:galaxy_selection}).
Other differences refer to tiles, the combined product of deep stacks, which we do not use.
The $u$-, $g$-, $r$-, $i$- and $z$-band data
were treated in exactly the same way as described in Paper~I.

\subsection{Accounting for foreground Milky Way reddening}
\label{accounting_foreground_mw_reddening}

\begin{figure}
\centering
\includegraphics[width=\columnwidth]{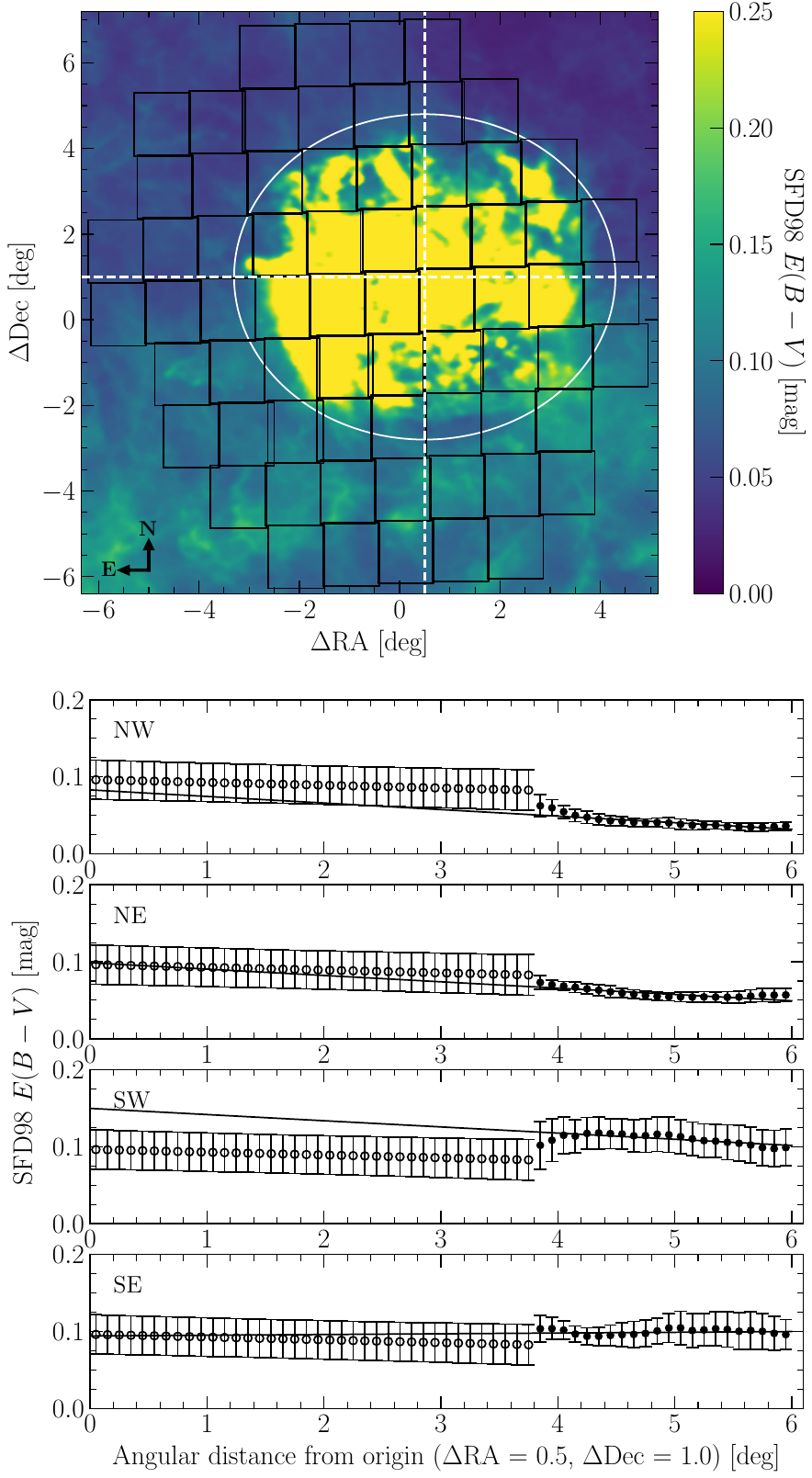} \caption[]{\emph{Top panel}: \cite{Schlegel98} foreground MW and residual
intrinsic LMC reddening map. The black polygons represent the positions of the VMC tiles covering the LMC.
The dashed white lines separate the LMC into different quadrants. The white circle (centred at
$\Delta \mathrm{RA}=0.5$, $\Delta \mathrm{Dec}=1.0$\,deg and with a radius of 3.8\,deg) marks the region within which the SFD98 $E(B-V)$
values are unreliable. \emph{Bottom panel}: SFD98 $E(B-V)$ values in each quadrant as a function of angular distance. The solid points denote the median $E(B-V)$ value calculated from the
SFD98 reddening map between distances of 3.8 and 6\,deg in steps of 0.1\,deg and the associated error bars represent the standard
deviation in that annulus. The linear best fit to the solid points in each quadrant is also shown as the straight black line. The open 
points are the same in each panel and denote the median of the four fits in steps of 0.1\,deg covering the region within which the SFD98
$E(B-V)$ values are unreliable. The error bars signify the standard deviation between the four fits.}
\label{fig:schlegel_red_map}
\end{figure}

To produce a dust map of the LMC we need to take into account the reddening along the line-of sight. 
In Papers~I and II we adopted a rather rudimentary method to account for the foreground MW reddening towards the SMC by 
adopting a mean $E(B-V)$ value for all objects based on the \citet*[hereafter referred to as SFD98]{Schlegel98} reddening map 
purposefully neglecting the main body and Wing of the SMC (see fig. 4 of Paper~I). Whilst such a treatment, to
zeroth order, should adequately reflect the ``average'' foreground MW reddening, it is likely to hide fluctuations or trends in
$E(B-V)$ across the SMC. In light of this, we hereby adopt a revised method that we describe below.

The top panel of Fig.~\ref{fig:schlegel_red_map} shows the SFD98 foreground MW and residual intrinsic LMC reddening map with the positions of the VMC tiles covering the LMC overlaid\footnote{We use the \texttt{Python} module
\textsc{sfdmap} (\url{https://github.com/kbarbary/sfdmap}) to determine the SFD98 $E(B-V)$ value at a given position.}. The inner
circle (centred at $\Delta \mathrm{RA}=0.5$, $\Delta \mathrm{Dec}=1.0$\,deg and with a radius of 3.8\,deg) marks the region within
which the SFD98 $E(B-V)$ values are unreliable due to insufficiently resolved temperature structures. Outside of this inner circle
the SFD98 $E(B-V)$ values are believed to be reliable (see e.g. \citealp{Schlegel98,Skowron21}) and clearly show a significant amount
of substructure in the high-latitude MW cirrus (this difference is especially noticeable in regions to the south and to the north of
the LMC). For objects farther than 3.8\,deg from the centre of the inner circle we simply de-redden the fluxes provided by
\textsc{lambdar} using \textsc{sfdmap}. For objects within the inner circle, however, we construct a radial $E(B-V)$ profile from
linear extrapolations of fits to regions outside of the inner circle.

The bottom panel of Fig.~\ref{fig:schlegel_red_map} shows the SFD98 $E(B-V)$ values as a function of angular distance from
$\Delta \mathrm{RA}=0.5$, $\Delta \mathrm{Dec}=1.0$\,deg. The solid points denote the median $E(B-V)$ value calculated in each
quadrant between the two white circles in steps of 0.1\,deg using \textsc{sfdmap} and the associated error bars represent the
standard deviation in that annulus. The best-fitting linear curve to the solid points in each quadrant is also shown as the straight
black line. The open points are the same in each panel and denote the median of the four fits in steps of 0.1\,deg covering the
region within which the SFD98 $E(B-V)$ values are unreliable. The error bars signify the standard deviation among the four fits.
These open points represent the radial $E(B-V)$ profile that we adopt to de-redden the fluxes for objects that lie within the inner
circle i.e. objects within this region are de-reddened according to their angular distance from
$\Delta \mathrm{RA}=0.5$, $\Delta \mathrm{Dec}=1.0$\,deg. From Fig.~\ref{fig:schlegel_red_map} it is clear that there are enhanced 
levels of reddening in the south-west quadrant with respect to the other three quadrants, and this likely accounts for the observed
offset between the extrapolated line and the calculated radial profile within 3.8\,deg in this quadrant which is not observed in the
other quadrants. At the time of submission of our work to the journal, \cite{Chen2022} presented a dust map of the MW derived from the SED of stellar sources within 5 kpc. They combined astrometry from {\it Gaia} with optical and near-IR photometry from various large-scale surveys towards the Magellanic Clouds. Their MW reddening map towards the LMC shows an average $E(B-V)$ of 0.06 mag with values as high as 0.15 in the south of the disc (an area outside our inner circle). The small inhomogeneities and clumps in the inner regions do not appear to exceed this value (their Fig.~10). Our extrapolation corresponds to an average $E(B-V)$ of about 0.1 mag (Fig.~\ref{fig:schlegel_red_map}) which is in line with the maximum values derived by \cite{Chen2022}, but may overestimates by a few percent the extinction towards some specific lines of sights.

The extinction coefficients adopted to convert a given $E(B-V)$ value into an extinction in a specific bandpass are
those listed in eq.~1 of Paper~I and should be used in conjunction with the unscaled SFD98 $E(B-V)$ values as they already take into
account the \cite{Schlafly11} recalibration. After de-reddening the fluxes and retaining only objects for which \textsc{lambdar}
measures positive fluxes in at least four out of the eight available bandpasses, our full LMC sample contains 2,474,235 objects.

\subsection{Fitting the SEDs with \sc{lephare}}
\label{fittings_seds_lephare}

Papers~I and II provide a comprehensive overview of fitting SEDs using
\textsc{lephare}\footnote{\url{http://cesam.lam.fr/lephare/lephare}}. However, we briefly discuss the notable difference in this work
with respect to the earlier works in this series. In Papers~I and II we primarily included quasi-stellar object (QSO)/AGN templates (hereafter collectively referred to as AGN) as the level of AGN contamination in our sample was unclear. With
no comprehensive catalogue of AGN behind the SMC [except for the areas covered by, e.g, the XMM--\emph{Newton} survey \citep{Haberl12} as
well as incomplete spectroscopic samples (e.g. \citealp*{Kozlowski11}; \citealp{Kozlowski13})] we included such templates to not only
account for the presence of AGN, but also to provide more flexibility with regards to fitting the SEDs. The inclusion of the
AGN templates in \textsc{lephare} likely resulted in degeneracies in the best-fitting templates as the number of objects classified as
AGN in the full SMC sample was markedly higher than those \textsc{lephare} classified as galaxies (see table~2 of Paper~II). This,
in addition to the previously adopted criteria for selecting background galaxies from the deep VMC PSF photometric catalogues (see
Section~\ref{selection_of_background_galaxies}), likely contributed to the low number of ETGs in the central regions of the SMC.

\begin{figure}
\centering
\includegraphics[width=\columnwidth]{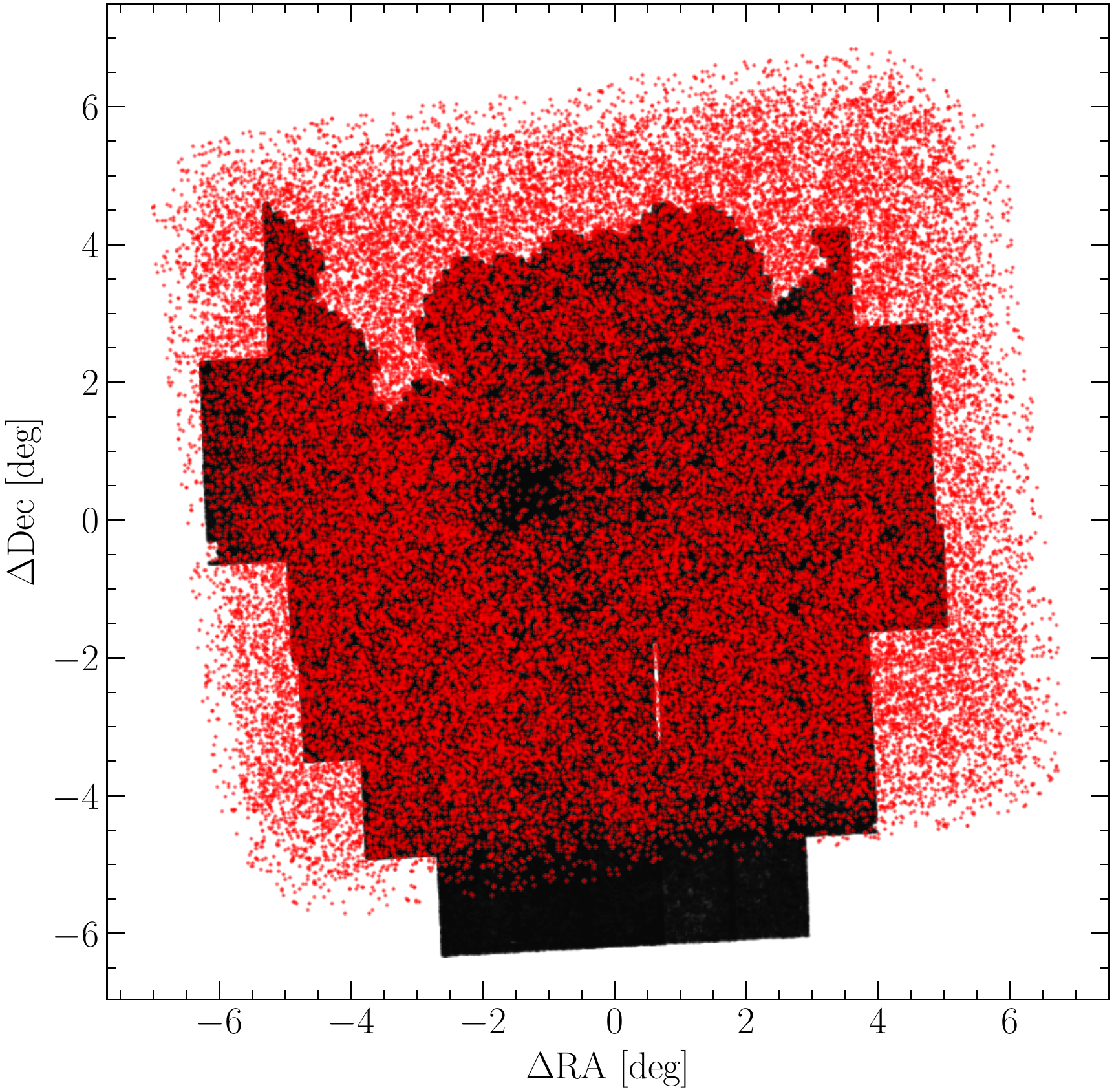} \caption[]{Positions of the 54,612 radio sources detected in the
888\,MHz ASKAP EMU data towards the LMC (red points; \citealp{Pennock21}) overlaid on the full SMASH-VMC sample (black points).}
\label{fig:radio_gal_overlap}
\end{figure}

In this study we make use of the recently published catalogue of radio-detected sources in the direction of the LMC by \cite{Pennock21}
based on 888\,MHz Australian Square Kilometre Array Pathfinder (ASKAP) Evolutionary Map of the Universe (EMU) radio continuum survey
data. These data cover $\simeq$\,120\,deg$^{2}$ and the catalogue contains 54,612 sources separated into {\sc{Gold}}, {\sc{Silver}}
and {\sc{Bronze}} categories (see section~3.2 of \citealp{Pennock21} for details). Although the catalogue also contains both MW sources
in the foreground of the LMC as well as sources in the LMC itself, the vast majority of the radio sources are background objects, and
in particular, AGN. Fig.~\ref{fig:radio_gal_overlap} shows that the radio sources almost completely
cover the full LMC sample, except for a significant fraction of the lowest row of VMC tiles (LMC 2\_3--LMC 2\_7). We can therefore use this catalogue to systematically and effectively remove AGN contaminants -- as well as foreground interlopers -- from our sample and ensure that we are not reliant upon \textsc{lephare} for such discrimination.

There are two primary reasons for separating the AGN from the full LMC sample. First, as demonstrated in Paper~I, it is difficult to use
AGN to create a reddening map due to the varying levels of intrinsic reddening. By removing AGN from the full LMC sample, they are
unable to be misclassified by \textsc{lephare} and incorrectly incorporated into the creation of the reddening map. Second, the vast 
majority of spectroscopic redshifts for objects behind the LMC are associated with AGN. To test the validity of the \textsc{lephare}
photometric redshifts it is therefore best to fit such objects with more representative AGN templates so as to minimise any potential
biases introduced by using less well-suited galaxy templates. Following \cite{Pennock21}, we adopt a cross-match radius of 5\,arcsec
and find a total of 21,706 radio sources in the full LMC sample. Whilst the \cite{Pennock21} catalogue is complete down to 0.5\,mJy
across the field and contains the majority of the spectroscopically confirmed AGN in the combined SMASH--VMC footprint of the LMC, we
still identified an additional 122 AGN with spectroscopic redshift determinations from v7.2 of the Milliquas (Million Quasars)
catalogue\footnote{\url{http://quasars.org/milliquas.htm}} \citep{Flesch21} that were not present and include these to create our final
AGN sample comprising 21,828 sources. These sources are subsequently removed from the full LMC sample, reducing its number to 2,474,235.
Note that although the radio catalogue of sources in the direction of the LMC provided by \cite{Pennock21} is the most sensitive to date,
only $\sim$\,10--20 per cent of all AGN are radio-loud (see e.g. \citealp{Ivezic02,Jiang07}). We are therefore no doubt significantly
underestimating the level of AGN contamination in our sample of extragalactic objects behind the LMC, although the EMU end-of-survey
sensitivity should detect radio-quiet AGN out to $z \simeq 2$ \citep{Norris11}.

We run \textsc{lephare} as described in Paper~II on both the full LMC and AGN samples. We only use the empirical AVEROI\_NEW galaxy templates for the LMC sample in addition to the stellar templates, whereas for the AGN sample we replace the galaxy templates with the 10 AGN templates described in
\citet[see section~3.1 of Paper~I for details regarding these template libraries]{Polletta07}. As in Paper~II, we allow additional
intrinsic reddening [in terms of $E(B-V)$] to vary and allow this
additional reddening for all galaxy/AGN types. Typically, additional reddening is only used for later-type galaxies (Sc and bluer/later;
see e.g. \citealp{Arnouts07}) and AGN (see e.g. \citealp{Salvato09}) on the basis that the adopted templates may not fully account for
the widely varying levels of dust observed in such galaxies/AGN. For ETGs, and particularly in the optical/near-IR regime we are
interested in here, the range in dust is not as extreme and so the assumption is that the adopted templates fully account for their
dust contents and thus additional reddening is not required. Our method of using ETGs as reddening probes essentially builds upon
this assumption by suggesting that any additional reddening required by ETGs is due to a foreground extinguishing medium, in our case
the LMC. Tables~\ref{tab:lephare_output_gal} and \ref{tab:lephare_output_agn} provide the \textsc{lephare} outputs 
for the 2,474,235 and 21,828 sources in the full LMC and AGN samples,
respectively. The associated $\chi^{2}$ values for the best-fitting templates imply that 61 per cent of the full LMC sample are galaxies
(39 per cent stars) and 87 per cent of the AGN sample are AGN (13 per cent stars). Of the 61 per cent of objects classified as galaxies
in the full LMC sample, 15 per cent (222,752) are classified as ETGs.

\begin{table*}
\caption[]{A sample of the \textsc{lephare} output for the 2,474,235 sources in the full LMC sample. We only show the ID, RA and Dec.
(J2000.0), the best-fitting photometric redshift with associated $1\sigma$ limits, the maximum likelihood photometric redshift with
associated $1\sigma$ limits, the best-fitting galaxy template, the associated $\chi^{2}$ value for the best-fitting galaxy template
and the best-fitting $E(B-V)$ value. The full table available as Supporting Information contains other parameters.}
\begin{tabular}{l c c c c c c c c c c c}
\hline
ID   &   RA (J2000.0)   &   Dec. (J2000.0)   &   $z_{\rm{BEST}}$   &   $z_{\rm{BEST}}^{-1\sigma}$   &   $z_{\rm{BEST}}^{+1\sigma}$   &
$z_{\rm{ML}}$   &   $z_{\rm{ML}}^{-1\sigma}$   &   $z_{\rm{ML}}^{+1\sigma}$   &   Template$_{\rm{G}}^{a}$   &   $\chi^{2}_{\rm{G}}$   &
$E(B-V)$\\
   &   deg   &   deg   &   &   &   &   &   &   &   &   &   mag\\
\hline
1   &   73.77041   &   $-75.72749$   &   0.4906   &   0.4645   &   0.5143   &   0.4694   &   0.3974   &   0.5147   &   54   &   2.1045   &   0.05\\
2   &   73.82095   &   $-75.72746$   &   0.3292   &   0.3059   &   1.3738   &   1.1080   &   0.3329   &   1.5322   &   59   &   8.6474   &   0.00\\
3   &   73.78706   &   $-75.72451$   &   0.0200   &   0.0200   &   0.0277   &   0.0128   &   0.0041   &   0.0310   &   49   &   71.1816   &   0.15\\
4   &   73.74649   &   $-75.71641$   &   1.0280   &   0.7738   &   1.2179   &   0.9209   &   0.5290   &   1.1316   &   50   &   2.6083   &   0.15\\
5   &   73.50257   &   $-75.71626$   &   0.5961   &   0.5347   &   1.1331   &   0.9244   &   0.5870   &   1.2578   &   57   &   3.1876   &   0.50\\
6   &   73.83151   &   $-75.71592$   &   0.4862   &   0.4433   &   0.5389   &   0.4802   &   0.3976   &   0.5374   &   38   &   5.5756   &   0.00\\
7   &   73.48358   &   $-75.71535$   &   1.3214   &   1.3109   &   1.3322   &   1.3297   &   1.3041   &   1.3635   &   62   &   15.4637   &   0.40\\
8   &   73.77131   &   $-75.71414$   &   0.0555   &   0.0423   &   0.0667   &   0.0726   &   0.0423   &   2.8935   &   62   &   15.1095   &   0.05\\
9   &   73.76825   &   $-75.71168$   &  0.9478   &   0.7439   &   1.2849   &   0.9306   &   0.5771   &   1.1749   &   62   &   4.7682   &   0.30\\
10   &   73.82471   &   $-75.71136$   &   0.5050   &   0.4176   &   0.8201   &   0.5114   &   0.3058   &   0.7615   &   62   &   2.0733   &   0.10\\
\hline
\end{tabular}
\vspace{1pt}
\begin{flushleft}
$^{a}$ Best-fitting galaxy templates are as follows: (1--21) E, (22--37) Sbc, (38--48) Scd, (49--58) Irr, (59--62) Starburst.\\
$^{b}$ Context is a numerical representation in \textsc{lephare} specifying the combination of bands present in the input catalogue and
is defined as $\sum_{i=1}^{i=N} 2^{i-1}$, where $i$ is the band number (in our case $u=1$, $g=2$, ..., $J=7$, and $K_{\rm{s}}=8$), and
$N$ is the total number of bands.
\end{flushleft}
\label{tab:lephare_output_gal}
\end{table*}

\begin{table*}
\caption[]{Same as Table~\ref{tab:lephare_output_gal}, but for the 21,828 sources in the AGN sample.}
\begin{tabular}{l c c c c c c c c c c c}
\hline
ID   &   RA (J2000.0)   &   Dec. (J2000.0)   &   $z_{\rm{BEST}}$   &   $z_{\rm{BEST}}^{-1\sigma}$   &   $z_{\rm{BEST}}^{+1\sigma}$   &
$z_{\rm{ML}}$   &   $z_{\rm{ML}}^{-1\sigma}$   &   $z_{\rm{ML}}^{+1\sigma}$   &   Template$_{\rm{Q}}^{a}$   &   $\chi^{2}_{\rm{Q}}$   &
$E(B-V)$\\
   &   deg   &   deg   &   &   &   &   &   &   &   &   &   mag\\
\hline
1   &   74.13216   &   $-74.39312$   &   1.1025   &   1.0778   &   1.1794   &   1.1353   &   1.0840   &   1.1957   &   6   &   10.2503   &   0.45\\
2   &   73.66724   &   $-74.35413$   &   0.7129   &   0.5553   &   0.9163   &   0.8019   &   0.6263   &   1.0059   &   4   &   3.3765   &   0.00\\
3   &   74.62965   &   $-74.33360$   &   1.0313   &   0.7321   &   1.4235   &   1.0588   &   0.6518   &   1.5054   &   5   &   7.6673   &   0.15\\
4   &   73.07838   &   $-74.29228$   &   0.3555   &   0.3540   &   0.3623   &   0.3559   &   0.3414   &   0.3722   &   10   &   121.5800   &   0.15\\
5   &   70.73015   &   $-74.10165$   &   1.0855   &   1.0737   &   1.2082   &   1.1423   &   1.0872   &   1.2030   &   6   &   10.8134   &   0.10\\
6   &   70.58080   &   $-74.08259$   &   1.0907   &   1.0138   &   1.1312   &   0.8248   &   0.6040   &   1.0886   &   3   &   8.9963   &   0.00\\
7   &   76.32935   &   $-74.59744$   &   1.0284   &   1.0141   &   1.0474   &   1.0428   &   1.0138   &   1.1032   &   3   &   33.3055   &   0.35\\
8   &   76.59007   &   $-74.57986$   &   1.2702   &   1.0548   &   1.2807   &   1.2378   &   1.0576   &   1.2750   &   10   &   9.9513   &   0.00\\
9   &   75.99906   &   $-74.47083$   &   1.0737   &   1.0768   &   1.0807   &   1.0792   &   1.0655   &   1.0933   &   6   &   15.1357   &   0.40\\
10   &   77.26805   &   $-74.42954$   &   0.3326   &   0.3187   &   0.3475   &   0.3332   &   0.3125   &   0.3541   &   1   &   52.6541   &   0.05\\
\hline
\end{tabular}
\vspace{1pt}
\begin{flushleft}
$^{a}$ Best-fitting AGN templates are as follows: (1) Seyfert 1.8, (2) Seyfert 2, (3--5) type-1 QSO, (6--7) type-2 QSO, (8--9)
Starburst/ULIRG, (10) Starburst/Seyfert 2.\\
$^{b}$ Context is a numerical representation in \textsc{lephare} specifying the combination of bands present in the input catalogue and
is defined as $\sum_{i=1}^{i=N} 2^{i-1}$, where $i$ is the band number (in our case $u=1$, $g=2$, ..., $J=7$, and $K_{\rm{s}}=8$), and
$N$ is the total number of bands.
\end{flushleft}
\label{tab:lephare_output_agn}
\end{table*}

\begin{figure}
\centering
\includegraphics[width=\columnwidth]{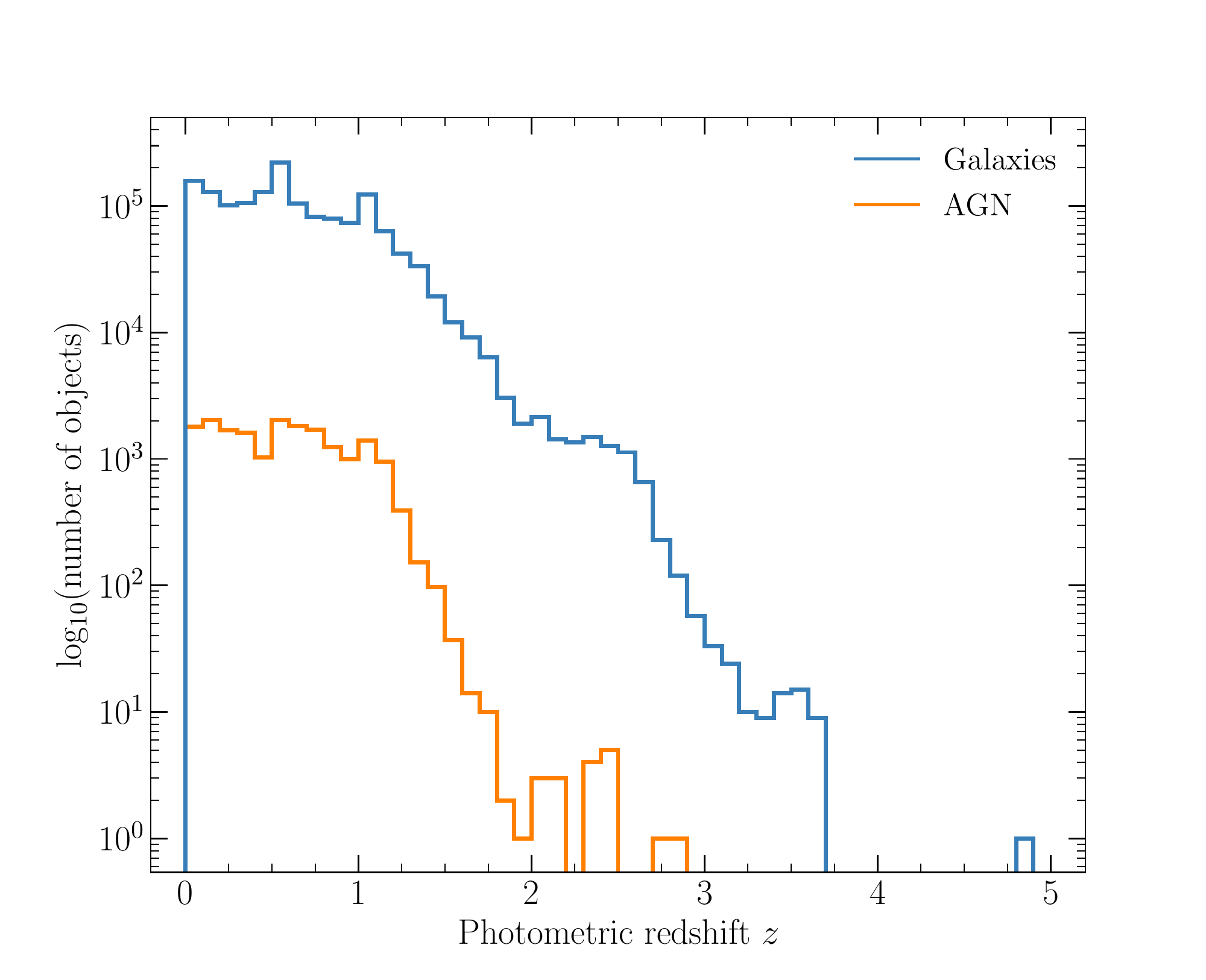} \caption[]{Distribution of redshifts for galaxies (blue) and AGN (orange) classified
by \textsc{lephare}.}
\label{fig:photo_z_dist_lmc_agn}
\end{figure}

Fig.~\ref{fig:photo_z_dist_lmc_agn} shows the photometric redshift distributions for the 1,504,987 objects classified as galaxies and the 19,017 objects classified as AGN in the full LMC and AGN samples, respectively. As discussed in section~2.3 of Paper~II, we choose 
the photometric redshift from the median of the maximum likelihood distribution ($z_{\rm{ML}}$) or if this
is not available\footnote{$z_{\rm{ML}}$ is only unavailable in cases of extremely poor fits ($\chi^{2} \gtrsim 1500$) and this occurs in about 
1 per cent of the galaxies.} the one with the minimum $\chi^{2}$ value ($z_{\rm{BEST}}$). The median redshift of both samples is very similar ($z_{\rm{med}}=0.56$
and 0.57 for the galaxy and AGN samples, respectively). Although this median redshift is almost identical to that of the objects
classified as galaxies behind the SMC in Paper~II, we refrain from making a direct comparison between the two galaxy populations due
to the combined effects of the different methods of selecting background galaxies as well as the use of different template sets between
the two studies. Any conclusions drawn from such a comparison (e.g. the different distribution of spectral types) should be treated with
caution and not used to argue for systematic differences between the two galaxy populations. Although such a comparison would be
worthwhile, this would necessitate a complete redetermination and analysis of the SMC sample that is beyond the scope of this study.

\subsubsection{Effects of missing $u$-band data}
\label{missing_u_band_data}

\begin{figure}
\centering
\includegraphics[width=\columnwidth]{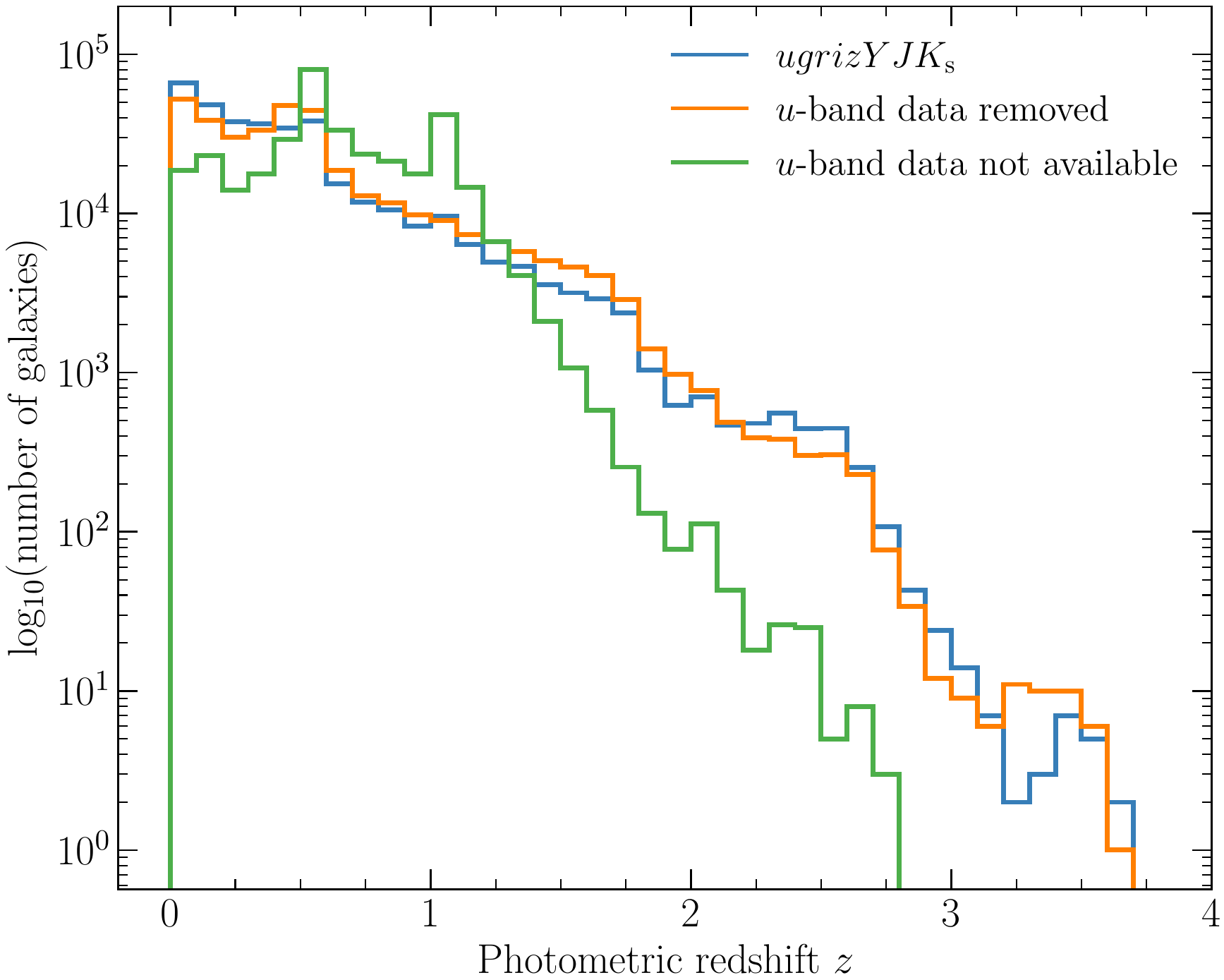} \caption[]{Distributions of galaxies redshifts in the full LMC sample. The blue line denotes galaxies for which the SED comprises all
eight available bands (from $u$ to $K_{\rm{s}}$). The orange line represents the same sample of galaxies as shown by the blue line, however we
have removed the measured $u$-band fluxes and re-run through \textsc{lephare}. The green line illustrates galaxies for which there
are no $u$-band data available but have measured fluxes in the seven remaining bands (from $g$ to $K_{\rm{s}}$).}
\label{fig:photo_z_dist_lmc_u_no_u}
\end{figure}

It is worth noting that we have not yet discussed the effects of non-complete $u$-band coverage across the combined SMASH--VMC footprint
of the LMC. From Fig.~\ref{fig:smash_vmc_coverage} it is evident that only the central regions of the LMC have $u$-band data from SMASH,
as well as a few isolated regions in the outskirts. In Fig.~\ref{fig:photo_z_dist_lmc_u_no_u} we show the photometric redshift
distributions for objects classified as galaxies in the full LMC sample. There are three samples of galaxies shown in
Fig.~\ref{fig:photo_z_dist_lmc_u_no_u}: i) objects that have measured fluxes in all eight available bands (from $u$ to $K_{\rm{s}}$), ii) as i) but with the $u$-band fluxes removed and re-run through \textsc{lephare} and iii) objects for which no $u$-band data are 
available, but which have measured fluxes in the remaining seven bands (from $g$ to $K_{\rm{s}}$). The number of galaxies in each sample is
limited to 350,000 to better illustrate systematic differences between the samples. The effect of simply removing the $u$-band data
(where available) is small but noticeable, in the sense that there are some variations in the redshift distributions across the whole
redshift range, but overall the distributions resulting from samples (i) and (ii) are broadly similar in terms of median redshift and shape (cf. $z_{\rm{med}}=0.36$ for the 8-band SED
sample and $z_{\rm{med}}=0.44$ for the reprocessed sample after removing the $u$-band data). The resulting samples of ETGs are also similar.
In contrast, the redshift distribution
for objects for which no $u$-band data are available is markedly different, both in terms of median redshift ($z_{\rm{med}}=0.59$) and as regards 
the apparent dearth of galaxies at higher redshifts. These differences likely stem from the SMASH observing strategy for which only regions with full $ugriz$ coverage have deep exposures, whereas the regions with only $griz$ coverage are covered by short
exposures (see Section~\ref{creating_fitting_seds_of_galaxies} for details). The latter sample of ETGs will be similar to that of ETGs within the same magnitude ranges resulting from deeper observations.
This further reinforces the importance of the $u$ band as
a powerful diagnostic to discriminate between low ($z \lesssim 0.5$) and higher redshift galaxies (see e.g. 
\citealp{Bellagamba12,Bisigello16}) as well as the need for homogeneous coverage (in terms of photometric bandpasses and exposure time)
across the survey area. 

\subsubsection{Comparison of spectroscopic and photometric redshifts for AGN behind the LMC}
\label{comparison_agn_behind_lmc}

\begin{figure}
\centering
\includegraphics[width=\columnwidth]{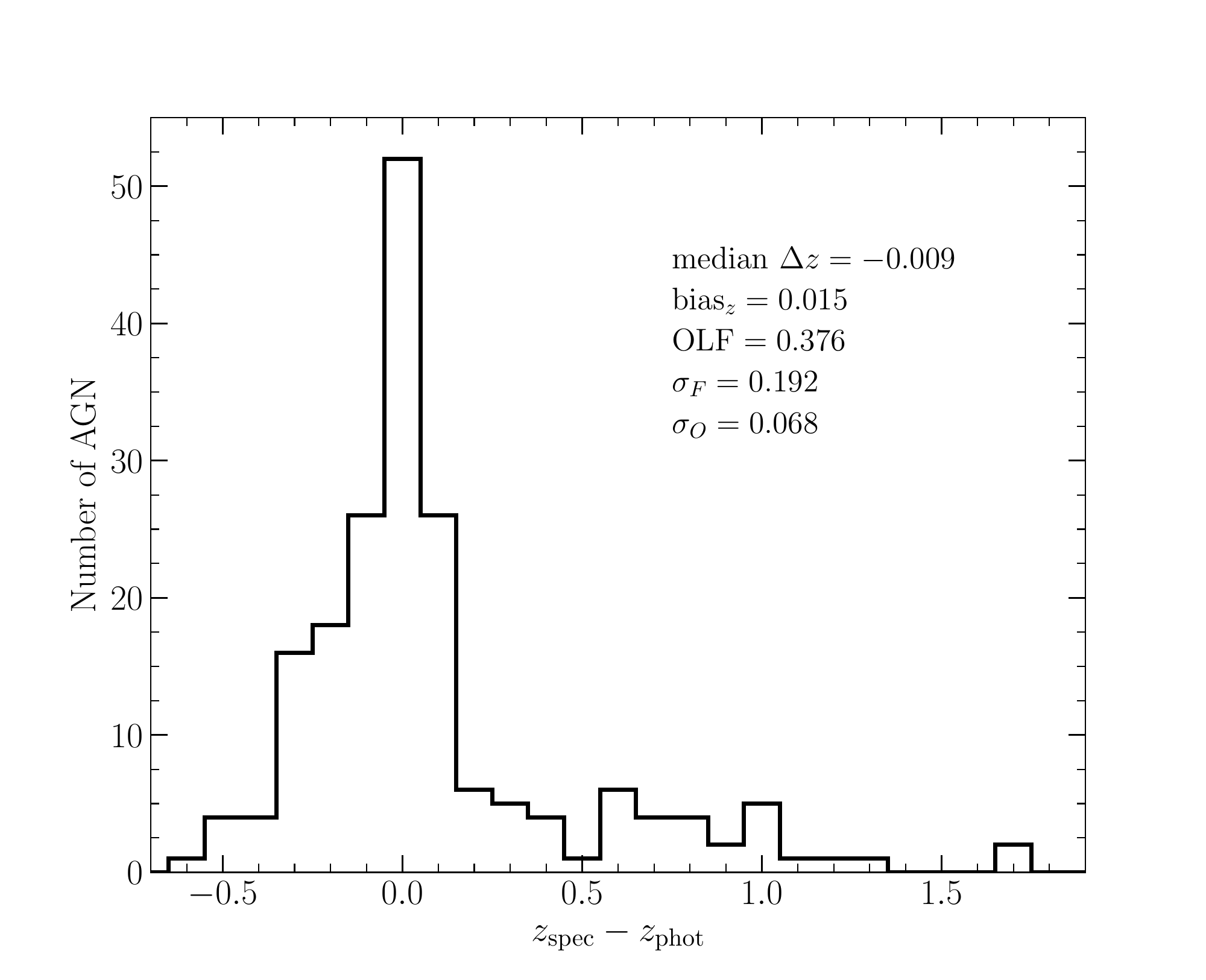} \caption[]{Histogram showing the differences between the
spectroscopically determined redshifts of AGN behind the LMC and the corresponding photometric redshifts calculated by \textsc{lephare}.
The various statistics regarding the comparison of the two redshift determinations are discussed in the text.}
\label{fig:z_spec_photo_agn_comp}
\end{figure}

\begin{table}
\caption[]{Spectroscopically determined and photometric redshifts for AGN behind the LMC. The full
table is available as Supporting Information.}
\begin{tabular}{l c c c}
\hline
Name   &   $z_{\mathrm{spec}}$   &   $z_{\mathrm{phot}}$   &   Ref.\\
\hline
PMN J0601--7238                     &   0.001   &   $0.014^{+0.031}_{-0.009}$   &   1\\
6dF J050434.2--734927	        &   0.045   &   $0.186^{+0.120}_{-0.121}$   &   2\\
MQS J053242.46--692612.2     &   0.059   &   $0.058^{+0.041}_{-0.050}$   &   3\\
MQS J043522.69--690352.9$^{a}$      &   0.061   &   $0.096^{+0.004}_{+0.004}$   &   3\\
PGC 3095709                            &   0.064    &   $0.158^{+0.068}_{-0.092}$   &   4\\
J050550.3--675017                     &   0.070   &   $0.155^{+0.071}_{-0.052}$   &   5\\
1E 0534--6740$^{a}$                  &   0.072  &   $0.234^{+0.006}_{+0.006}$   &   6\\
PGC 88452                                &   0.075    &   $0.086^{+0.024}_{-0.037}$   &   7\\
MQS J045554.57--691725.6      &   0.084    &   $0.005\pm0.003$   &   3\\
MQS J043632.30--704238.1      &   0.142    &   $0.182^{+0.207}_{-0.116}$   &   3\\
\hline
\end{tabular}
\vspace{1pt}
\begin{flushleft}
Notes: $^{a}$Associated 1$\sigma$ limits on $z_{\mathrm{BEST}}$ are unphysical (see section~2.3 and footnote 4 of Paper~II for details).\\
$^{b}$Associated $\chi^{2}$ value for best-fitting stellar template is lower than $\chi^{2}$ value for best-fitting AGN template.\\
References: (1) \protect\cite{Ajello20}, (2) \protect\cite{Jones09}, (3) \protect\cite{Kozlowski13}, (4) \protect\cite{Cowley84},
(5) \protect\cite{Dobrzycki05}, (6) \protect\cite{Crampton97}, (7) \protect\cite{Cristiani84}, (8)  \protect\cite{Kozlowski12},
(9)  \protect\cite{Wang91}, (10) \protect\cite{Dobrzycki02}, (11) \protect\cite{Tinney99}, (12) \protect\cite{Geha03},
(13) \protect\cite{Ivanov16}, (14) \protect\cite{Kostrzewa-Rutkowska18}.
\end{flushleft}
\label{tab:spec_phot_z_agn}
\end{table}

As an additional test we compare the derived photometric redshifts to spectroscopic redshifts of 189 AGN. Table~\ref{tab:spec_phot_z_agn} lists the comparison between the values and Fig.~\ref{fig:z_spec_photo_agn_comp} highlights a median difference of $\Delta z_{\rm{med}}=-0.009$, where $\Delta z_{\rm{med}}=z_{\rm{spec}}-z_{\rm{phot}}$, and median absolute deviation of $\Delta z_{\rm{MAD}}=0.191$. Note that of the 189 AGN in our sample,
five (MQS J050010.83--700028.5, 1E 0547--6745, MQS J054400.43--705846.2, MQS J043200.60--693846.5 and MQS J052528.91--700448.6) are
classified as stars by \textsc{lephare}. The statistics reported here, with the exception of the bias, outlier fraction (OLF), are very similar to those reported in
Paper~II for the sample of 46 AGN behind the SMC. From Fig.~\ref{fig:z_spec_photo_agn_comp} it is clear that there is an extended tail
towards higher values i.e. the spectroscopic redshift is larger than the photometric redshift and that this is more prevalent at
higher redshifts (see the full version of Table~\ref{tab:spec_phot_z_agn}). To test whether this tail is a consequence of our choice
of AGN templates, we re-fitted the SEDs using the AGN templates of \cite{Salvato09}. The tail persists even though these templates also include AGN-host
galaxy type hybrids (\citealp{Salvato09}). Note that studies have found that
template-based redshift estimates for AGN can be underestimated and that Gaussian and/or hierarchical Bayesian process photometric
redshift estimates perform significantly better at $z>1$ (see e.g. \citealp{Duncan18a,Duncan18b}).
Another potential cause for the underestimated redshifts could be contamination from neighbouring objects. Of the 28 AGN for which
$|z_{\rm{spec}}-z_{\rm{phot}}| > 0.5$, 21 (75 per cent) have neighbouring objects within 2\,arcsec that likely affect the
deblended SED of the AGN.

\section{Internal reddening of the LMC}
\label{intrinsic_reddening_lmc}

\begin{figure}
\centering
\includegraphics[width=\columnwidth]{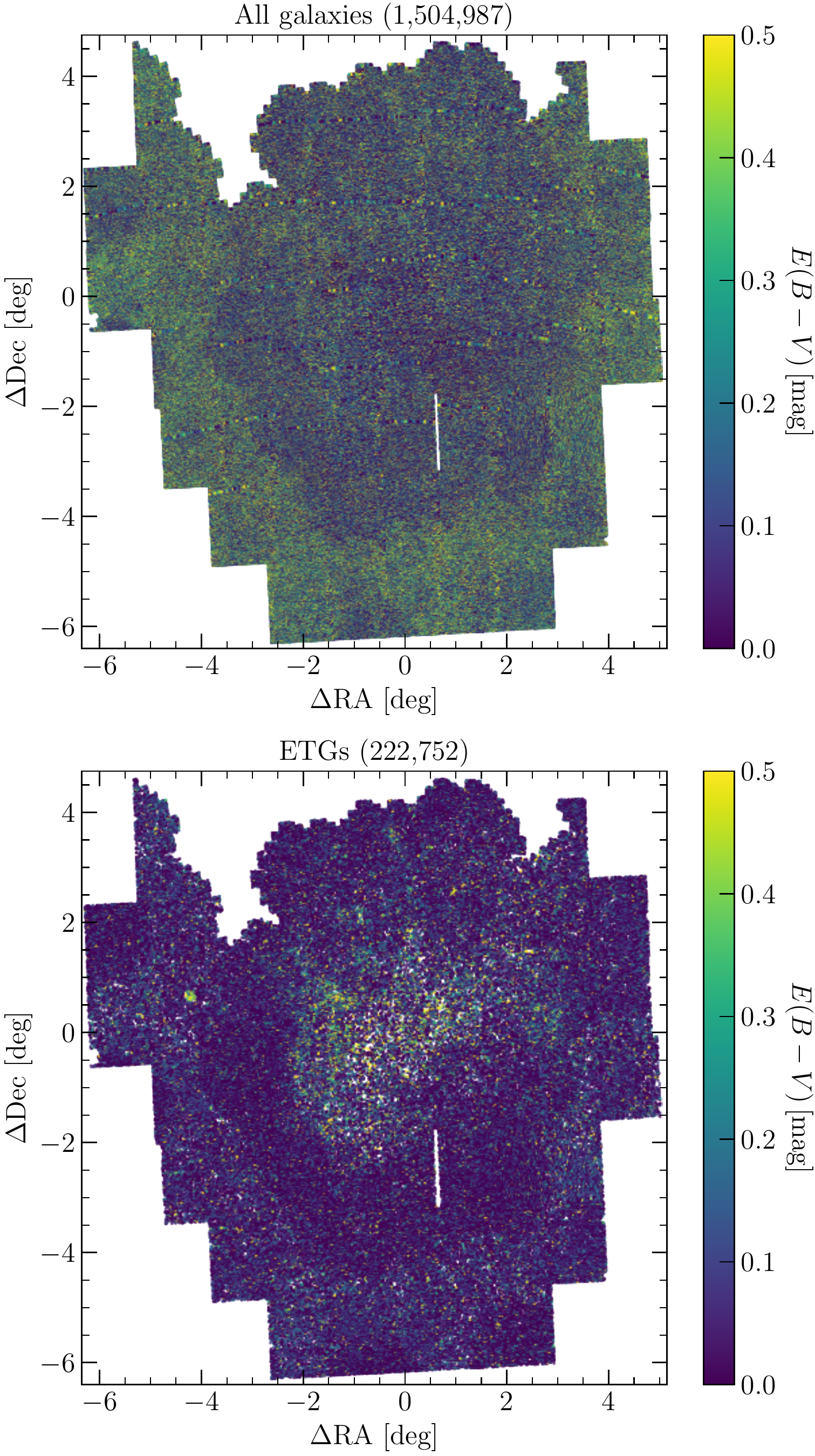} \caption[]{\emph{Top panel}: Reddening map covering the
combined SMASH--VMC footprint created using all objects classified as galaxies by \textsc{lephare}. \emph{Bottom panel}: As the top
panel, but using only objects classified as ETGs.}
\label{fig:red_map_LMC_full_indv_points}
\end{figure}

Fig.~\ref{fig:red_map_LMC_full_indv_points} shows the reddening map covering $\simeq$\,90\,deg$^{2}$ of the LMC based on the 1,504,987
objects classified as galaxies by \textsc{lephare} in the full LMC sample (top panel) in addition to the map created using only the
subsample of 222,752 objects classified as ETGs (bottom panel). From Fig.~\ref{fig:red_map_LMC_full_indv_points} there are two striking
observations. First, it is clear that the use of all galaxies (as discussed in Papers~I and II) effectively masks the regions where
significant amounts of dust intrinsic to the LMC are expected (e.g. along the bar region). This is primarily due to the inclusion of
late-type galaxies which are much more numerous than the ETGs and for which the best-fitting template requires additional reddening
(including lines of sight through the LMC that do not exhibit significant amounts of reddening), resulting in the
systematically higher reddening values across the LMC in the top panel of Fig.~\ref{fig:red_map_LMC_full_indv_points} compared to the
bottom panel. It is only by creating the reddening map using ETGs (with low levels of intrinsic dust themselves) that regions of high
internal reddening in the LMC become apparent. Second, there is an obvious difference between the regions that include $u$-band data
and those that do not (see also Section~\ref{missing_u_band_data}). Comparing Figs.~\ref{fig:smash_vmc_coverage} and 
\ref{fig:red_map_LMC_full_indv_points} it is evident that the regions for which $u$-band data are available tend to exhibit lower levels
of reddening (except for the central regions) than regions for which these data are not available. The enhanced reddening surrounding the galaxy beyond $3-4$ deg from the centre might therefore not be real. 
The MW correction is applied directly to the LAMBDAR fluxes prior to the SED-fitting (see Sec. 2.3). As we only allow the additional reddening to vary from $0.0-0.5$ in $E(B-V)$, there are no ETG galaxies for which the intrinsic reddening is negative.

\begin{figure*}
\centering
\includegraphics[width=\textwidth]{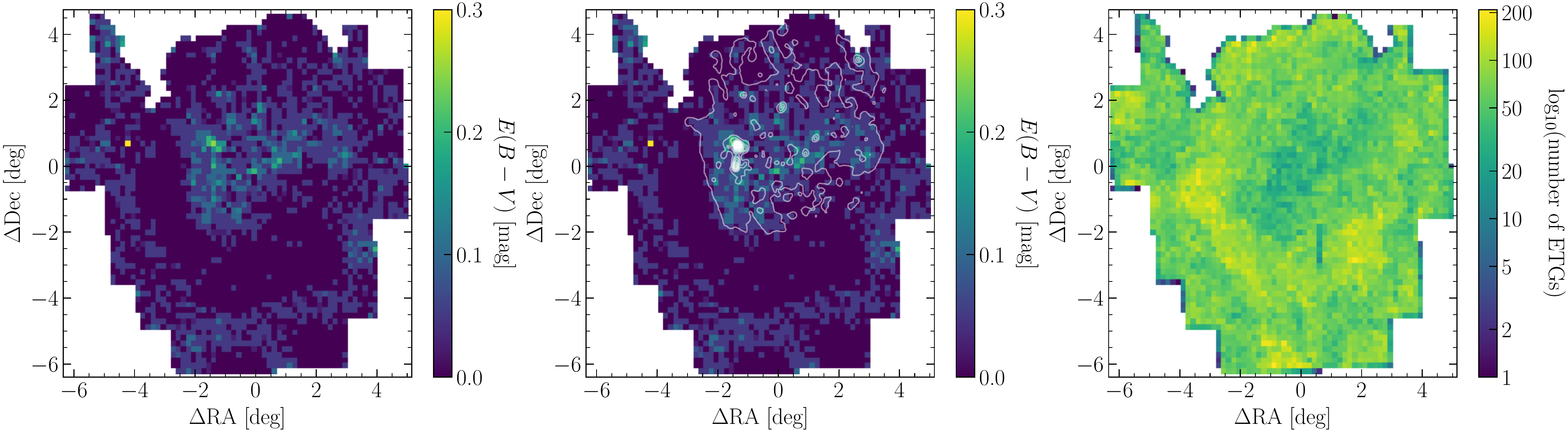} \caption[]{\emph{Left panel}: $10 \times 10$\,arcmin$^{2}$
resolution reddening LMC map using objects classified
as ETGs. \emph{Middle panel}: As the left panel, but with the IRAS 100\,$\mu$m dust emission contours overlaid to aid the reader in
terms of where enhanced levels of reddening (due to the presence of dust) are expected across the LMC. \emph{Right panel}: Number density of ETGs.}
\label{fig:red_map_LMC_full_10arcmin}
\end{figure*}

Fig.~\ref{fig:red_map_LMC_full_10arcmin} shows a $10 \times 10$\,arcmin$^{2}$ resolution reddening map using the same objects as
shown in the bottom panel of Fig.~\ref{fig:red_map_LMC_full_indv_points} to better illustrate large-scale patterns/features across
the LMC. Each bin corresponds to the median of the best-fitting $E(B-V)$ values and the median number of ETGs per bin is 63. 
Of 3348 bins covering the combined SMASH--VMC footprint of the LMC, only 52
(2 per cent) have fewer than 10 ETGs per bin and these are distributed around the outskirts, although there is an
obvious dearth of ETGs also associated with the vertical stripe shown in Fig.~\ref{fig:red_map_LMC_full_indv_points} resulting from
missing VMC data (see the right-hand panel of Fig.~\ref{fig:red_map_LMC_full_10arcmin}). It is also clear from
Fig.~\ref{fig:red_map_LMC_full_10arcmin} that there are fewer ETGs in the central regions of the LMC, which are more heavily affected
by crowding. Across the central regions of the LMC we find that the minimum number of ETGs in any given bin is 18, which although
significantly lower than the median of 63 across all bins, does not appear to be systematically affected [in terms of the $E(B-V)$ value]
with respect to the median $E(B-V)$ values in the neighbouring bins (for which the number of ETGs ranges from 20 to 82). However, there might be ETGs missing due to high extinction regions, which will introduce a bias towards lower extinctions, since both crowding and extinction increase inwards.

Also apparent from Fig.~\ref{fig:red_map_LMC_full_10arcmin} is the excellent agreement between the regions of enhanced reddening across
the central regions of the LMC and the regions that exhibit emission at longer wavelengths due to the
presence of dust (as traced by the IRAS 100\,$\mu$m emission contours). The regions of the LMC for which SMASH $u$-band data are available
completely cover these regions and so the transition from regions where we
expect dust to be present to regions where little/no dust is present is reflected in the reddening values determined from the ETGs. Farther
from the LMC centre, where no $u$-band data are available (except for some small isolated regions; see Fig.~\ref{fig:smash_vmc_coverage})
enhanced levels of reddening again become noticeable. Given the lack of $u$-band data in these regions, and the aforementioned agreement 
between the ETG-determined regions of enhanced reddening and far-IR emission, one should treat the absolute reddening values in these
outer regions of the LMC with caution [although they are, to within the uncertainties on the median $E(B-V)$, also consistent with zero
intrinsic reddening].

The use of ETGs to trace the intrinsic reddening of the LMC has been shown to reproduce the pervasive enhanced levels of reddening across
the LMC bar region that we know are present from existing far-IR observations (see e.g. \citealp{Chastenet17} and references therein).
In addition, there are several isolated regions for which the median reddening is further enhanced, with respect to this low-level
pervasive enhancement. These enhancements coincide with well-known star-forming regions throughout the bar (e.g. N44, N105, N113, N120, 
N144, and N206, as well as the Tarantula Nebula and the molecular ridge south of 30 Doradus) as well as overdensities observed in morphology
maps based on young stars. Interestingly, highest median reddening [$E(B-V)=0.35$\,mag;
$\Delta$RA\,$\simeq$\,--4.2, $\Delta$Dec\,$\simeq$\,0.7\,deg] lies far from the LMC bar in a region that is not actively forming stars
and for which there are no hints of enhanced far-IR emissions in the IRAS 100\,$\mu$m data (see Fig.~\ref{fig:red_map_LMC_full_10arcmin}). 

\begin{figure}
\centering
\includegraphics[width=\columnwidth]{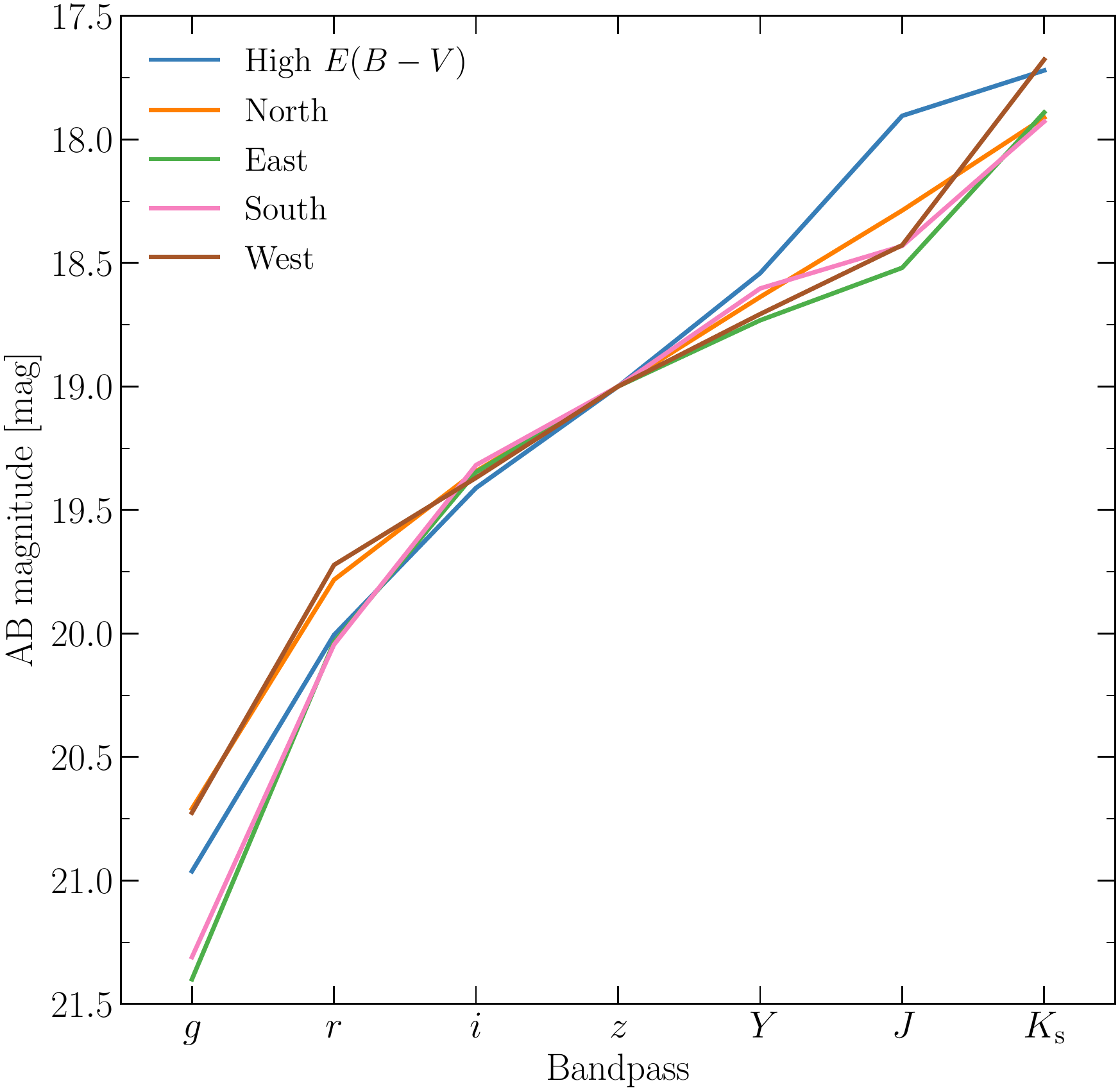} \caption[]{Comparison of median SEDs for all ETGs in the bin with the
highest intrinsic reddening (see text) as well as neighbouring bins to the north, south, east and west.}
\label{fig:etg_sed_comparison}
\end{figure}

It should be noted, that this region of the combined SMASH--VMC footprint lacks $u$-band data and so that could partly be responsible
for the anomalous reddening value. However, the lack of other such enhancements in regions that lack $u$-band data suggests this is
not the sole reason. Furthermore, the number of ETGs used to compute the median reddening is almost identical to the median across
the whole LMC (cf. 61 and 63) and there is a clear overdensity of ETGs with higher than average reddening values (see 
Fig.~\ref{fig:red_map_LMC_full_indv_points}). Fig.~\ref{fig:etg_sed_comparison} shows a comparison's between the median SED for all ETGs
in the bin with the highest intrinsic reddening and the ETGs in the neighbouring bins to the north, south, east and west. Note that for
comparison sake we have normalised all SEDs to a $z$-band magnitude of 19.0\,mag. Although there is very good agreement between the SEDs
across most of the optical/near-IR regime, there is a clear difference in the $J$-band regime, such that the median SED of the ETGs in the
high reddening bin is $\simeq$\,0.5\,mag brighter than those in the neighbouring bins. This difference is likely the reason behind the
anomalously high intrinsic reddening in this particular bin and may be related to a higher level of background substructure observed
in the VISTA $J$-band deep stack images covering this region of the sky.

The majority of the bins covering the combined SMASH--VMC footprint
have $E(B-V)=0.0$\,mag [$E(B-V)_{\rm{mean}}=0.06$\,mag] and a median uncertainty of $\sigma_{E(B-V)_{\rm{med}}}=0.09$\,mag.
For bins with only a single ETG, or for which all best-fitting reddening values are identical, we assume an uncertainty of 0.05\,mag corresponding to the spacing implemented in the SED process. Table~\ref{tab:lmc_red_bins} lists the median reddening values, standard deviations and the number of ETGs in each bin across the LMC footprint.

\begin{table*}
\caption[]{Reddening values derived from ETGs within 3348 bins across the LMC. The full table is available as Supporting Information.}
\begin{tabular}{c c c c c}
\hline
$\Delta$RA (J2000.0)   &   $\Delta$Dec. (J2000.0)   &   $E(B-V)_{\mathrm{med}}$   &   $\sigma_{E(B-V)_{\rm{med}}}$   &   No. ETGs\\
deg   &   deg   &   mag   &   mag   &\\
\hline
$-2.54675$   &   $-6.21777$   &   0.050   &   0.089   &   77\\
$-2.38055$   &   $-6.21777$   &   0.050   &   0.074   &   58\\
$-2.21435$   &   $-6.21777$   &   0.100   &   0.065   &   66\\
$-2.04816$   &   $-6.21777$   &   0.050   &   0.059   &   47\\
$-1.88196$   &   $-6.21777$   &   0.050   &   0.071   &   68\\
$-1.71576$   &   $-6.21777$   &   0.000   &   0.105   &   52\\
$-1.54957$   &   $-6.21777$   &   0.025   &   0.109   &   56\\
$-1.38337$   &   $-6.21777$   &   0.000   &   0.083   &   58\\
$-1.21718$   &   $-6.21777$   &   0.000   &   0.080   &   90\\
$-1.05098$   &   $-6.21777$   &   0.000   &   0.089   &   63\\
\hline
\end{tabular}
\label{tab:lmc_red_bins}
\end{table*}

\section{Discussion}
\label{discussion}

This study produced a map of the total intrinsic reddening of the LMC for a $\simeq$\,90\,deg$^{2}$ region covered by both the SMASH and VMC surveys. Using ETGs we have demonstrated the ability to identify regions with enhanced levels
of reddening that coincide with regions of enhanced far-IR emission. We proceed to compare our reddening map
with maps created from other tracers to determine how well the various tracers agree in terms of
both the line-of-sight reddening values as well as dust distribution morphology. As with the comparison in Paper~II, in the figures presented below we do not make any correction for the difference in the volume sampled between the reddening map based on ETGs and those based on various LMC stellar components. If necessary, we remove the foreground MW reddening from the literature maps following the same prescription as described in Section~\ref{accounting_foreground_mw_reddening}. The
SFD98 $E(B-V)$ values have been demonstrated to be overall systematically overestimated (see e.g. \citealp{Yuan13}) and so whilst their use to de-redden the \textsc{lambdar} fluxes was justified (as the extinction coefficients already accounted for this overestimation), their use to de-redden the reddening values in the literature maps would result in us effectively overestimating the foreground MW reddening, and as such adopt a scaling of 0.86 to reflect the recalibration of \cite{Schlafly11}.

\subsection{Literature LMC reddening maps}
\label{lmc_reddening_maps_literature}

\subsubsection{\protect\cite{Inno16}}
\label{inno16}

As part of our comparison in Paper~II we adopted the Cepheid sample of \cite{Joshi19}, however here we instead adopt that of
\cite{Inno16}. Although the reddening values are in very good agreement, the latter covers a larger area of the LMC disc and
is thus preferable as a comparison  data set. \cite{Inno16} studied the structure of the LMC disc using
multi-wavelength observations of Classical Cepheids (from the $V$-band in the optical to the $W1$-band in the mid-IR). As
part of this study, \cite{Inno16} used a multi-wavelength fitting of the reddening law to the apparent distance moduli
to determine the colour excess to individual Cepheids and thus create a reddening map covering $\simeq$\,80\,deg$^{2}$ of the
LMC disc. For more details regarding the methodology, readers are referred to the works of \cite*{Freedman91} and
\cite*{Gallart96} as early examples. \cite{Inno16} assumed a value of $R_{V} = 3.23$, where $R_{V}$ denotes the total to selective absorption, as part of their reddening analysis
and provide the individual reddening values in the form of $E(B-V)$. The highest median reddening corresponds to $0.65$ mag whereas the mean reddening corresponds to $0.02$ mag.

\subsubsection{\protect\cite{Choi18}}
\label{choi18}

\cite{Choi18} used $g$- and $i$-band data from SMASH to create a $2.67 \times 2.67$\,arcmin$^{2}$ resolution
reddening map covering $\simeq$\,165\,deg$^{2}$ of the LMC from a sample of $\sim$\,2.2 million RC stars.
As opposed to simply adopting a given intrinsic RC $g-i$ colour,
\cite{Choi18} instead measured the intrinsic colour radial profile using a ``clean'' RC sample based on
regions in the outskirts of the LMC that are essentially dust-free (see their figs. 4 and 7). The stars in these
regions were de-reddened using the SFD98 dust map before constructing the intrinsic colour radial profile.
The reddening map was created by comparing the observed $g-i$ colour map and the intrinsic $g-i$ colour map which
is based on the intrinsic colour radial profile. Note that the radial profile only covers an angular distance range
of 2.7 to 8.5\,deg (from the LMC centre) and thus \cite{Choi18} adopted the intrinsic RC colour at a distance of
2.7\,deg for the innermost regions ($< 2.7$\,deg) and similarly the intrinsic RC colour at a distance of 8.5\,deg
for the outermost regions ($8.5-10.5$\,deg). To convert the $E(g-i)$ values presented in the \cite{Choi18} reddening map,
we use the extinction coefficients listed in eq.~1 of Paper~I to yield $E(g-i) = 1.617 \times E(B-V)$. In the resulting map the highest median reddening is of $0.35$ mag and the mean reddening is of $0.02$ mag.

\subsubsection{\protect\cite{Skowron21}}
\label{skowron21}

\cite{Skowron21} used $V$- and $I$-band data from OGLE-IV (\citealp*{Udalski15}) to create a reddening map covering $\simeq$\,180\,deg$^{2}$ of the LMC using RC stars\footnote{\url{http://ogle.astrouw.edu.pl/cgi-ogle/get_ms_ext.py}}.
The resolution of the reddening map varies from $1.7\times1.7$\,arcmin$^{2}$ in the central regions
to $\simeq$\,$27\times27$\,arcmin$^{2}$ in the outer regions. \cite{Skowron21} determined an intrinsic RC $V-I$ colour radial profile by de-reddening the observed RC colours using the SFD98 dust map. Note that the authors exclude the central 4.1\,deg radius
region around the LMC centre due to the incorrect reddening values provided by the SFD98 dust maps as well as
some outer regions due to spurious RC colour measurements (see their section~4 for details). To estimate the intrinsic RC colour
in the inner and outer regions, \cite{Skowron21} used spectroscopic information from Apache Point Observatory Galactic Evolution Experiment (APOGEE) spectra of red giant
stars taken from \cite{Nidever20} to first calculate the LMC metallicity gradient and then a sample of intermediate-aged
(3--9.5\,Gyr) clusters to determine the variation of the intrinsic RC colour as a function of metallicity. These two
relations are then combined to provide an intrinsic colour change as a function of distance from the LMC centre of
--0.002\,mag\,deg$^{-1}$ that is used to calculate the intrinsic RC colour in the aforementioned excluded regions
(see their fig. 13). $E(V-I)$ values are converted into $E(B-V)$ using the relation from \cite*{Cardelli89}. We find a highest median reddening of $0.28$ mag and a mean reddening to $0.01$ mag.

\subsubsection{\protect\cite{Cusano21}}
\label{cusano21}

\cite{Cusano21} used a combination of near-IR time-series photometry from the VMC and optical light curves from OGLE-IV to
study the structure of the LMC as traced by $\simeq$\,29,000 RR Lyrae stars. As part of this analysis, they also determined optical $E(V-I)$
reddening values by comparing the observed colours of these stars to their intrinsic colour (based on an empirical relation
connecting the intrinsic colour of the star to its $V$-band amplitude and period; see also \citealp*{Piersimoni02}). Although
\cite{Cusano21} do not provide the individual $E(V-I)$ reddening values, they do provide the individual $K_{\rm{s}}$-band
extinction values (see their table~3) and so we convert these to $E(B-V)$ values using the extinction coefficient listed
in eq.~1 of Paper~I [$A_{K_{\rm{s}}} = 0.308 \times E(B-V)$]. The highest median reddening corresponds to $0.67$ mag and the mean reddenign to $0.02$ mag.

\subsubsection{\protect\cite{Utomo19}}
\label{utomo19}

\cite{Utomo19} homogeneously reanalysed archival far-IR maps from \emph{Herschel} in four Local Group galaxies (including the LMC at
$\sim$\,1\,arcmin resolution) by fitting the SED of dust from 100 to 500\,$\mu$m with a modified blackbody emission model (see
\citealp{Chiang18} for details regarding the model fitting). For our purposes, we are interested in the dust mass surface density,
$\Sigma_{\rm{d}}$, returned by the best-fitting model, as this can then be converted into $E(B-V)$ which can then be directly compared to the values we derive using ETGs. IR emission across these wavelengths primarily captures emission from relatively large dust grains that are in thermal equilibrium with the local radiation field, and represent the bulk of the total dust mass. The smaller grains however contribute to the reddening because of their surface to volume ratio.
We note that the dust map of the LMC adopted in this work (Chiang et al. priv. comm.) differs slightly from that calculated in \cite{Utomo19} in two small ways. First, whereas \cite{Utomo19} adopt a simple power law (see their eq.~5), the dust map included here instead uses a broken power law (see eqs.~2 and 3 of \citealp{Chiang21}) as this yields a better fit to the data (see \citealp{Chiang18}). Second, as a result of modifying the form of the fit, the emissivity at 160\,$\mu$m (the reference wavelength in \citealp{Utomo19} and which is determined by calibrating the models to the Milky Way cirrus where the dust mass is known, see e.g. \citealp{Jenkins09,Gordon14}) increases slightly to $\kappa_{160\,\mu\rm{m}}=20.73\pm0.97$\,cm$^{2}$\,g$^{-1}$ (cf. $18.7\pm0.6$\,cm$^{2}$\,g$^{-1}$ in \citealp{Utomo19}).

In Paper~II (see section~4.1.3) we demonstrated a method to convert $\Sigma_{\rm{d}}$ to $E(B-V)$ following the formalism of
\cite{Whittet03} and for which it was necessary to make assumptions regarding the values of the $V$-band extinction efficiency factor, $Q_{V}$, as well as the radius, $a$, and density, $\rho_{\rm{d}}$, of the dust grains. Here we adopt a simpler, albeit model-dependent, approach using the following conversion (see \citealp{Draine14} and references therein):

\begin{equation}
A_{V} = 0.7394 \left( \frac{\Sigma_{\rm{d}}}{10^{5}\,\mathrm{M_{\odot}/kpc^{2}}} \right) \,\mathrm{mag}
\end{equation}

\noindent and then convert $A_{V}$ into $E(B-V)$ as in \citep{Cardelli89}. 
Despite the different methods to convert $\Sigma_{\rm{d}}$ into $E(B-V)$, the median difference in the resulting $E(B-V)$ values is
only $\Delta E(B-V)_{\rm{med}}=0.004$\,mag. Furthermore, significant differences (greater than the 0.05\,mag level) affect only 1 per
cent of the total number of pixels covering the area of the LMC observed by \emph{Herschel} and coincide with the most intense regions of star formation. As in Paper~II, we do not need to subtract the foreground MW
contribution as this has already been accounted for as part of the SED modelling.

\subsubsection{Mazzi et al. (2021)}
\label{mazzi_2021}

\cite{Mazzi21} used near-IR VMC data to determine the SFH across $\simeq$\,96\,deg$^{2}$ of the
LMC with a spatial resolution of 0.125\,deg$^{2}$. The derivation of the SFH consists of determining the linear combination of partial
models that best fit the observed $K_{\rm{s}}, Y-K_{\rm{s}}$ and $K_{\rm{s}}, J-K_{\rm{s}}$ colour--magnitude Hess diagrams. These models
include the effects of extinction, and so by finding the best-fitting combination of models, \cite{Mazzi21} determined a
representative ``mean'' $V$-band extinction for each subregion. We transform $A_{V}$ into $E(B-V)$ following \cite{Cardelli89}
[$A_{V} = 3.1 \times E(B-V)$]. Note that we only adopt the $A_{V}$ values resulting from the $K_{\rm{s}}, J-K_{\rm{s}}$ fits due to
possible issues with the absolute $Y$-band calibration. Due to SFH analysis
being performed on discretised regions across the LMC, we are unable to compare this reddening map to the map based on ETGs at a
resolution of $10 \times 10$\,arcmin$^{2}$. To facilitate a comparison, we will degrade the resolution of the ETG reddening map and
discuss this map separately (see Section~\ref{comparing_different_reddening_tracers}).\\

\subsection{Comparing different reddening tracers}
\label{comparing_different_reddening_tracers}

\begin{figure*}
\centering
\includegraphics[width=\textwidth]{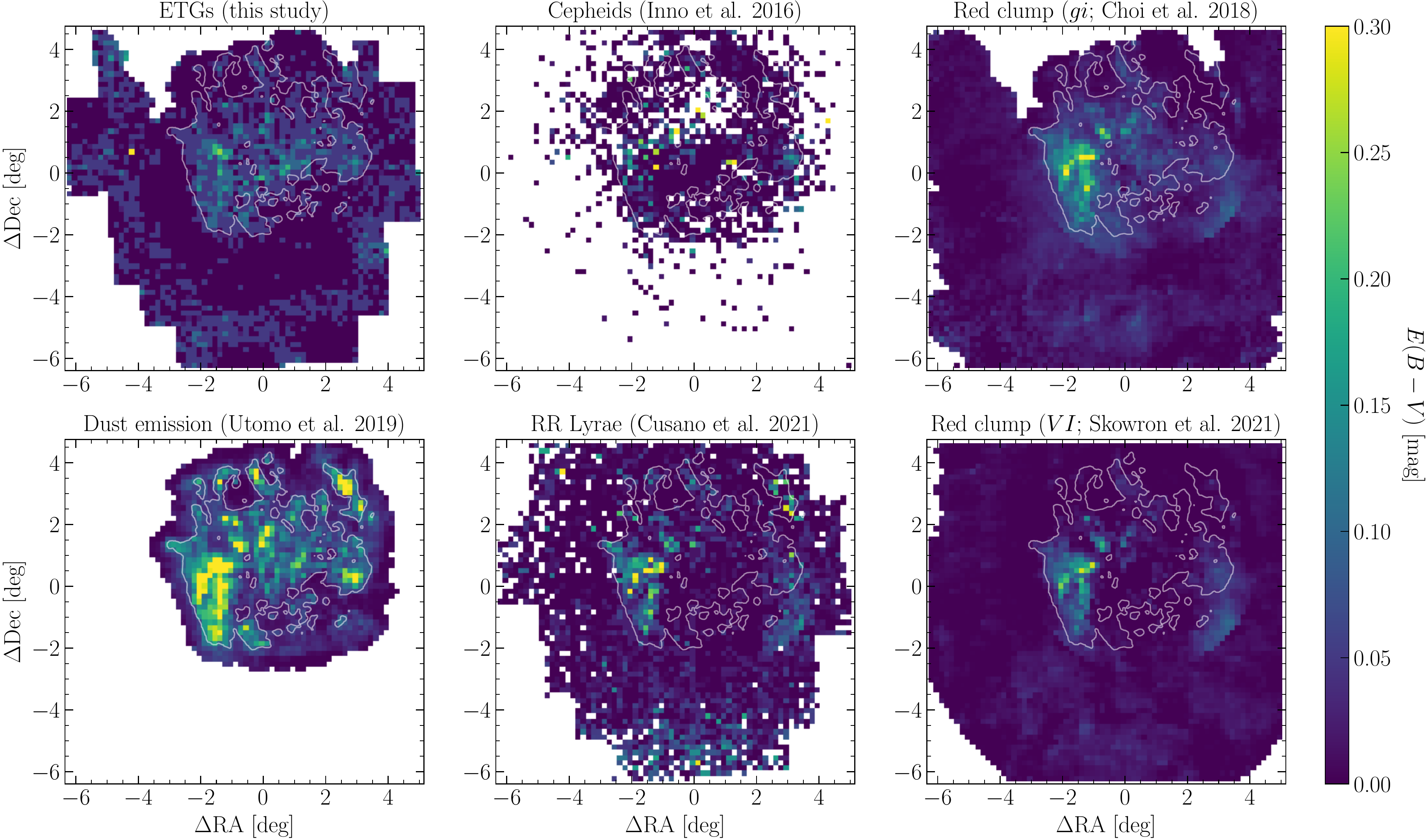} \caption[]{$10 \times 10$\,arcmin$^{2}$ resolution reddening
maps of the LMC. Each panel refers to a different literature source/tracer. The contour in each panel represents the low-level
pervasive dust associated with the central regions of the LMC as traced by the IRAS 100\,$\mu$m emission. Note that the colour bar range
has been limited to $0 \leq E(B-V) \leq 0.3$\,mag to better illustrate low-level features within the maps. Note also that the ETG extinctions should be roughly twice that of the Cepheids, RR Lyrae and RC values since the latter do not sample the full path through the LMC.}
\label{fig:red_map_comparison}
\end{figure*}

\begin{figure*}
    \centering
    \includegraphics[width=\textwidth]{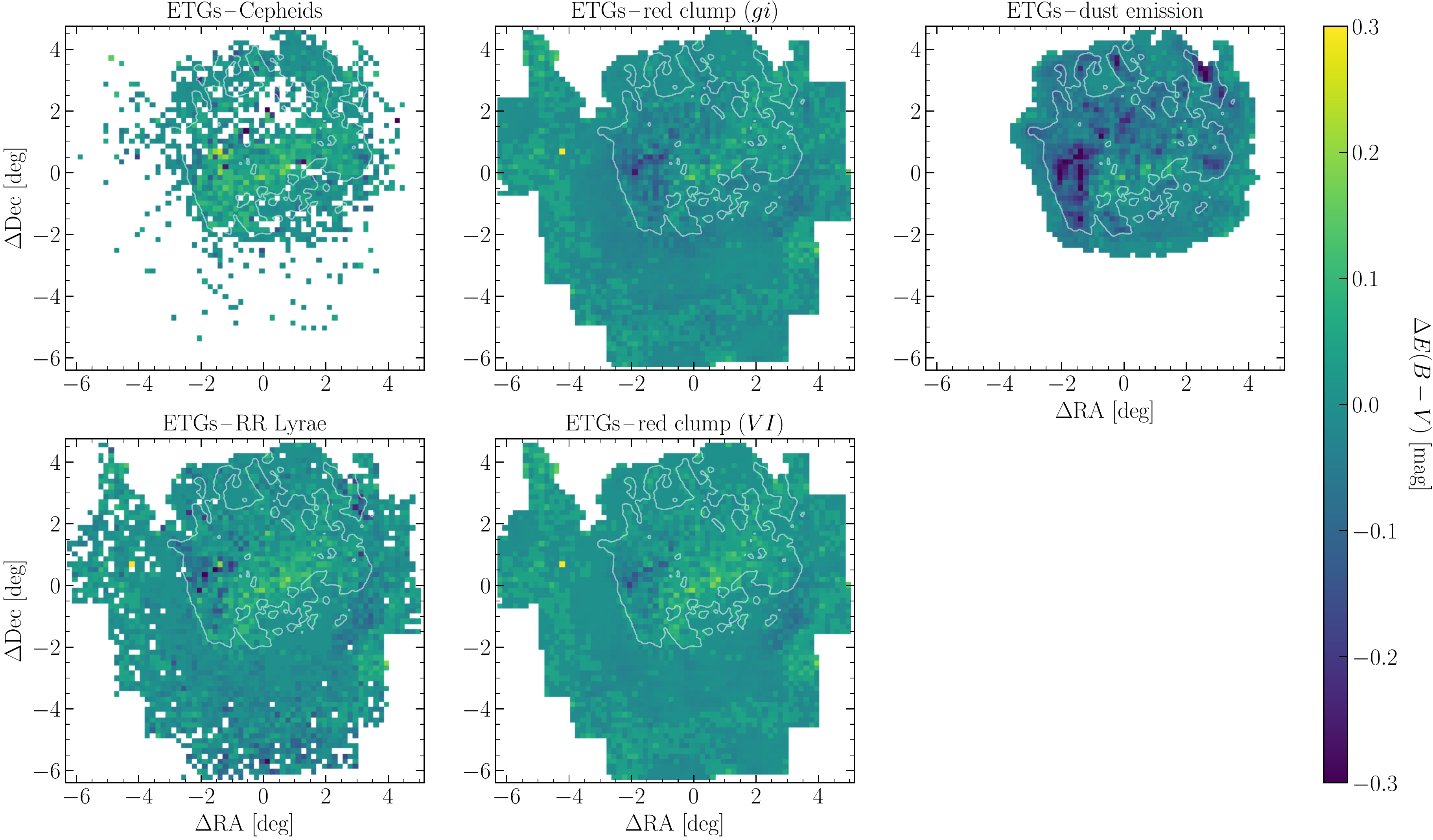}
    \caption[]{As Fig.~\ref{fig:red_map_comparison}, but showing a $10 \times 10$\,arcmin$^{2}$ resolution differential reddening maps of the LMC, such that $\Delta E(B-V) = E(B-V)_\mathrm{ETG} - E(B-V)_\mathrm{tracer}$. Note that the colour bar range has been limited to $-0.3\leq E(B-V)\leq 0.3$ mag to better illustrate low-level features within the maps.}
    \label{fig:red_map_difference}
\end{figure*}

Fig.~\ref{fig:red_map_comparison} shows the comparison between the LMC reddening map based on ETGs and the other maps resampled onto
the same $10 \times 10$\,arcmin$^{2}$ resolution. The majority of the maps used in our comparison have higher spatial resolutions than that shown in 
Fig.~\ref{fig:red_map_comparison} and so some small-scale features (of the order of tens of pc in size) will be masked. Despite this
resampling, any large-scale features present in the maps will be retained and evidenced in Fig.~\ref{fig:red_map_comparison}. In addition, Fig.~\ref{fig:red_map_difference} shows the difference between the reddening map based on ETGs and the literature maps such that $\Delta E(B-V)=E(B-V)_\mathrm{ETG}-E(B-V)_\mathrm{tracer}$.

Whilst the various reddening tracers shown in Fig.~\ref{fig:red_map_comparison} all exhibit enhanced levels of intrinsic reddening in some areas across the central regions of the LMC (in particular the Tarantula Nebula and the molecular ridge south of 30 Dor), not all are consistent with the morphology of the low-level pervasive dust emission as traced by the IRAS 100\,$\mu$m emission. In particular, we see clear differences between the stellar tracers (Cepheids, RR Lyrae and RC) and those that sample ETGs and far-IR emission. However, the dust emission or the Cepheids maps are likely very highly biased (the former by heating sources, the latter by being near regions of star formation) and therefore not appropriate to compare to the values presented here. The older stars are more likely to be representative of random lines of sight (modulo the factor of two difference).

Reddening values derived from pulsating stars (Cepheids and RR Lyrae stars) are lower in the central region of the galaxy (at the location of the bar) compared to regions east, west and north of it. These tracers follow a spatial distribution typical of young and old stars, respectively, with most of them located within a structure of a few kpc along the line of sight (e.g. \citealp{Inno16,Cusano21}). The lower level of reddening suggested by pulsating stars in the bar region compared to ETGs could be explained by the predominant location of these sources in front of the dust rather than behind it. This is however unlikely because both types of variables occupy the entire thickness of the galaxy. 
Whilst the sample of Cepheids and RR Lyrae stars are highly complete they suffer from incompleteness/blending, due to crowding, in the central regions (e.g. \citealp{Holl18}). This effect reduces the amplitude of variation with respect to that of isolated stars, and makes the stars redder which corresponds to a lower reddening towards them (see eqn. 2 of \citealp{Cusano21}).
The western edge of bar is consistently reddened in all maps, but the continuous extension towards the south is only prominent in the maps derived from RR Lyrae and RC stars. The latter shows also reddening in the central bar region, especially in the map from \cite{Choi18}. RC stars sample the same range of distances as RR Lyrae stars, but are significantly more numerous which may explain their capability to trace reddening across a larger volume. 

Similarly to our findings, the middle of the northern arm (a structure parallel to the bar) shows the largest reddening values compared to the other regions within it. Maps extending to the outer regions confirm enhanced reddening values in the south. In our map, the region of 30 Dor and of the molecular ridge south of it appear as reddened as the bar region. This may be a selective effect due to a lack of ETGs in regions of high extinction and/or crowding introducing a bias to lower reddening values (see also Sect. \ref{intrinsic_reddening_lmc}). Both effects do influence the molecular ridge region whereas crowding is the dominant effect in the bar region.

The differential reddening maps imply that, by sampling the full line of sight of the LMC, the intrinsic reddening values inferred from the background galaxies are on average higher than those inferred from stellar tracers. If we allow for a factor of 2 difference to account for this depth effect and consider that the median uncertainty on the reddening is of $\sigma_{E(B-V)_\mathrm{med}}=0.09$ mag, we find that our map is consistent with those from RC stars for which the highest median reddening differs by $0.19$ mag (\citealp{Skowron21}) and $0.26$ mag (\citealp{Choi18}), respectively. On the contrary, the highest median reddening for the dust emission exceeds that of our map by $0.50$ mag, for the Cepheids by $0.60$ mag and for the RR Lyrae stars by $0.62$ mag. Note that these values are not shown in Fig.~\ref{fig:red_map_difference} because the reddening scale has been limited to $\pm 0.3$ mag to better illustrate low-level features. The mean reddening differences are however consistent with zero for all tracers.

\begin{figure*}
\centering
\includegraphics[width=\textwidth]{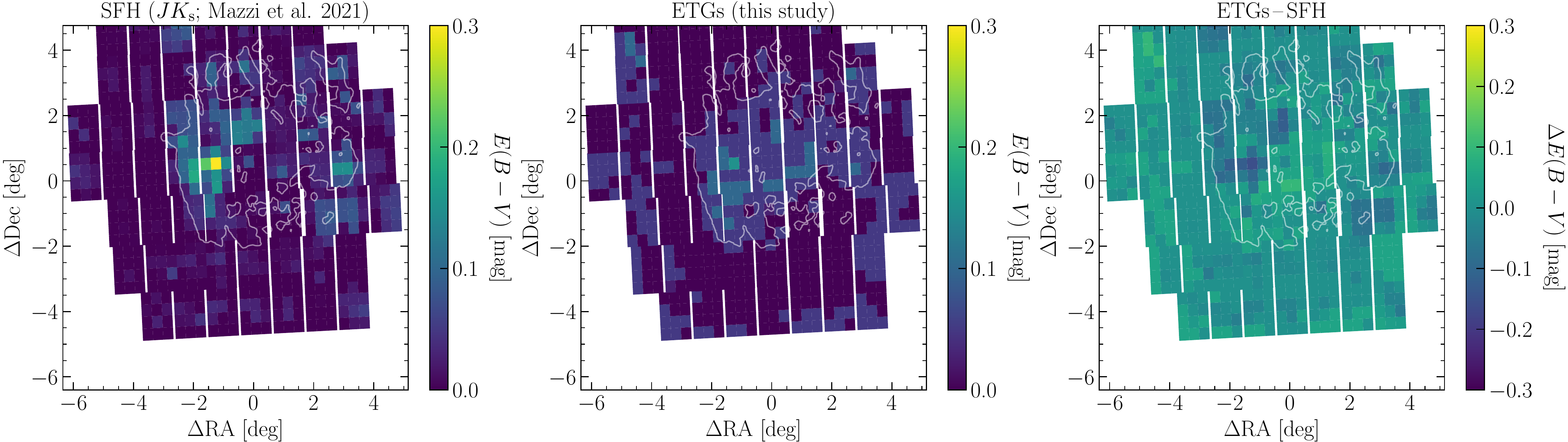}
\caption[]{Left panel: Reddening map of the LMC  derived from the SFH study by \cite{Mazzi21}. The spatial resolution corresponds to bins of $21.0 \times 21.5$\,arcmin$^{2}$ whereas 
the area shown is the same as in Figs. \ref{fig:red_map_LMC_full_indv_points}, \ref{fig:red_map_LMC_full_10arcmin} and \ref{fig:red_map_comparison}, hence the white space in the south and not all bins shown in the north. The contour represents the low-level pervasive dust associated with the central regions of the LMC as traced by the IRAS 100\,$\mu$m emission. Note that the colour bar range has been limited to $0 \leq E(B-V) \leq 0.3$\,mag to better illustrate low-level features within the maps. Middle panel: As the left panel, but showing the ETG reddening map adopting the spatial resolution of Mazzi et al.. Right panel: As the left panel, but showing the differential reddening map, such that $\Delta E(B-V) = E(B-V)_\mathrm{ETG} - E(B-V)_\mathrm{SFH}$. Note that the colour bar range has been limited to $-0.3\leq E(B-V)\leq 0.3$ mag to better illustrate low-level features within the map.}
\label{fig:sfh_map_comparison}
\end{figure*}

The comparison between our reddening map and the map derived from the SFH analysis is shown instead in Fig. \ref{fig:sfh_map_comparison}. In this case, the spatial resolution of each map corresponds to bins of $21.0 \times 21.5$\,arcmin$^{2}$. The SFH shows the highest reddening values in star forming regions (30 Dor, Constellation III and at the west end of the bar) which are not representative of the line-of-sights we trace in our study. Apart from other localised overdensities comparable to the size of a few bins the overall reddening is low. In particular, the central bar region appears as  extinct as the outer regions whereas in our study the bar and the overall inner region of the galaxy appear more extinct than its surroundings. By differentiating the SFH and ETG reddening maps we obtain a rather smooth distribution which agrees within the factor of 2 higher distance sampled by the ETGs. This suggests that a significant fraction of the dust is located beyond the bar sampled by the SFH analysis. The bar region is also affected by extreme crowding and in this region the SFH analysis relies more heavily on RC and RGB stars than on main-sequence turn-off and subgiant stars; these regions are associated to a low likelihood (\citealp{Mazzi21}).  

\begin{figure*}
\centering
\includegraphics[width=\textwidth]{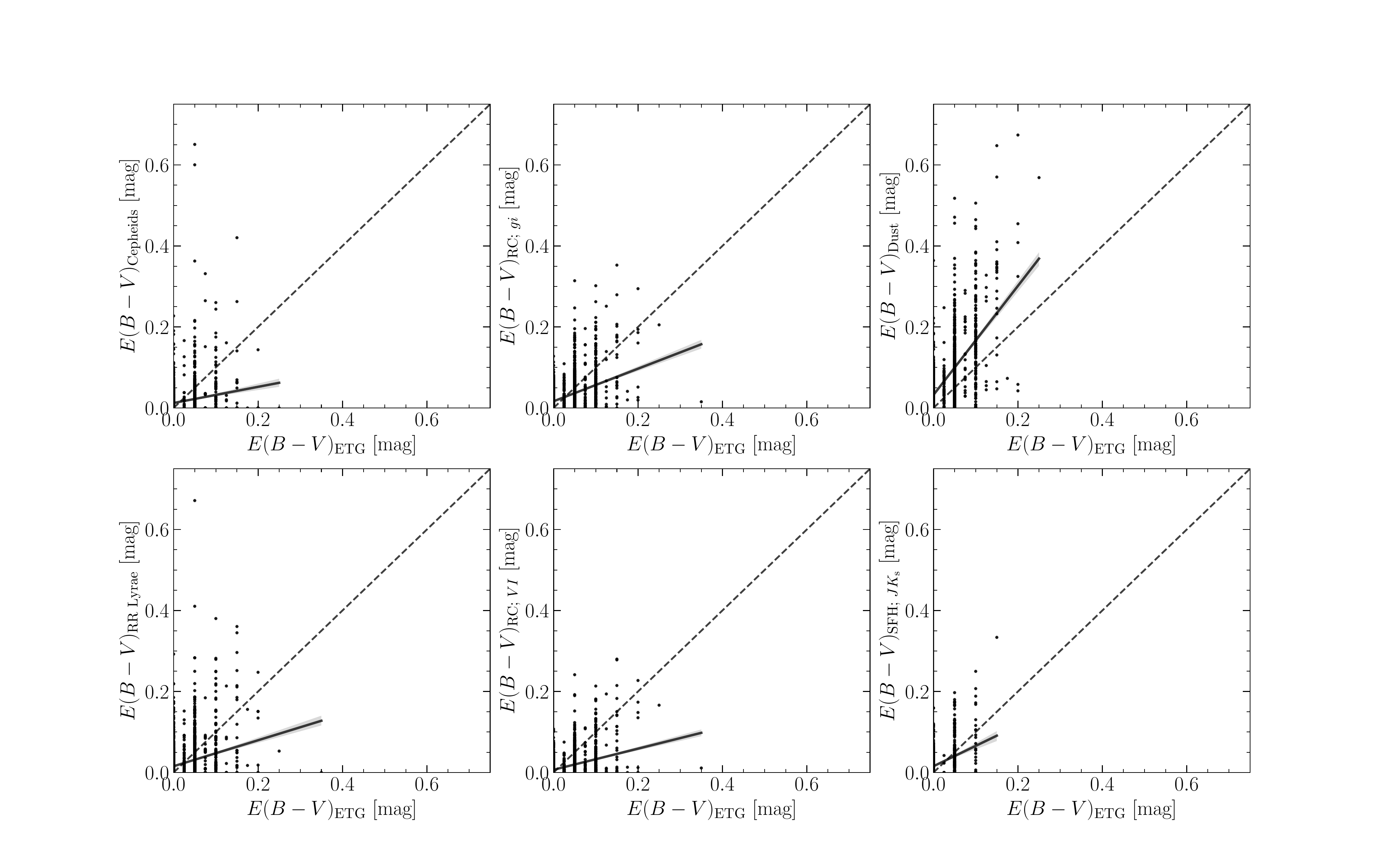} \caption[]{Comparison of the reddening values in the 10$\times$10 resolution ETG reddening map and those of the other reddening tracers. Note that the comparisons only include bins that cover the combined SMASH-VMC footprint of the LMC. The solid line and shaded region in each panel represent the linear best fit and corresponding 67 per cent confidence interval. The dashed line in each panel denotes the one-to-one correspondence between the two reddening tracers.}
\label{fig:scatterplot}
\end{figure*}

To further inspect possible correlations between the ETGs reddening and the reddening derived from the other tracers we plot in Fig.~\ref{fig:scatterplot} the relation between them. The best fit lines are clearly different from the one-to-one relation, except for the reddening derived from the dust emission which follows a similar trend, but is  offset by about 0.5 mag. We already mentioned earlier that despite the influence of heating sources there is a good agreement with the distribution of the low-level pervasive dust emission traced by the IRAS at 100 $\mu$m.

\begin{figure*}
\centering
\includegraphics[width=\textwidth]{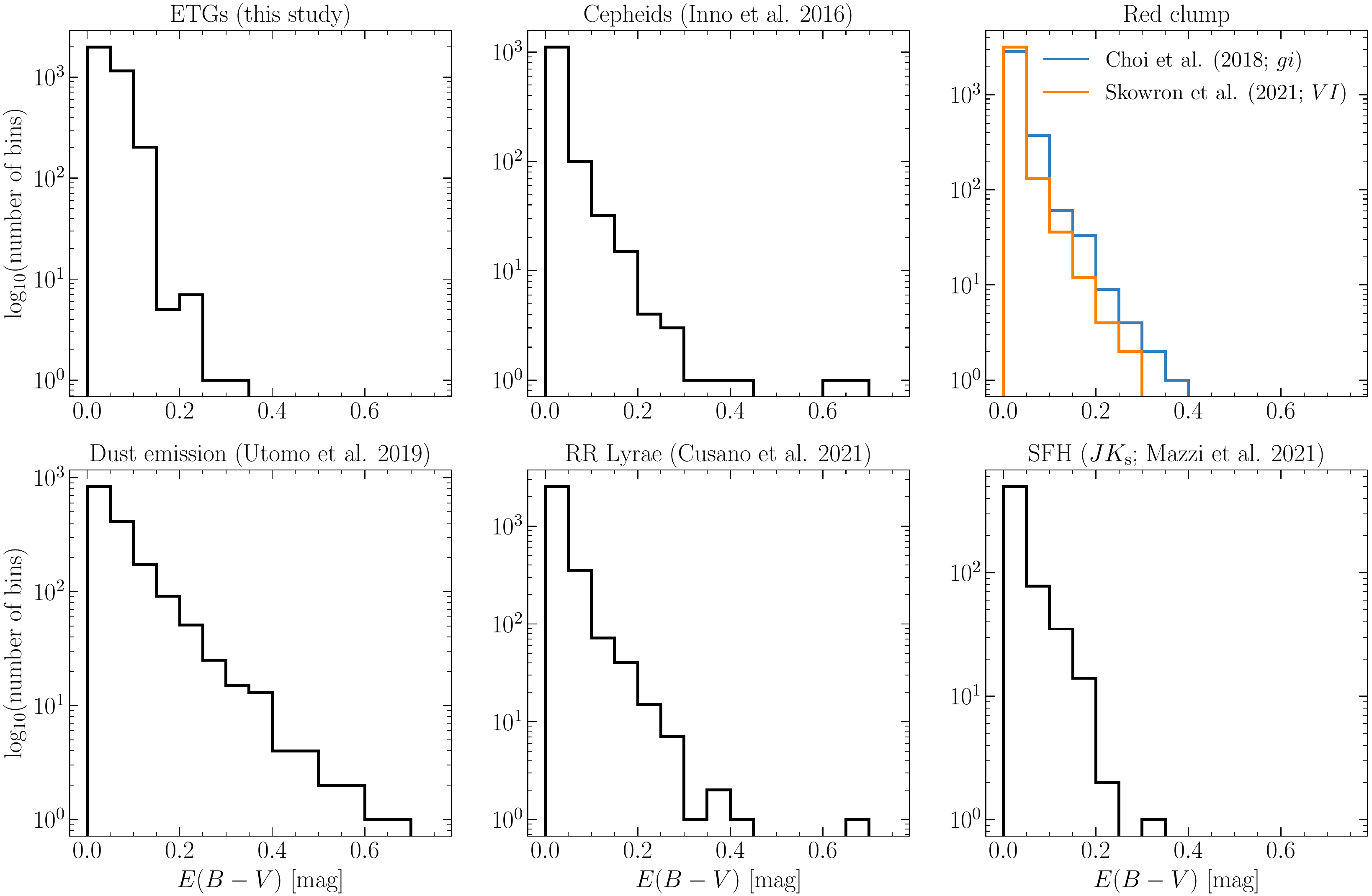} \caption[]{Histograms showing the distribution of reddening values in the 10$\times$10 resolution reddening maps shown in Figs.~\ref{fig:red_map_comparison} and \ref{fig:sfh_map_comparison}. Note that the histograms only include bins that cover the combined SMASH-VMC footprint of the LMC.}
\label{fig:histogram}
\end{figure*}

Figure \ref{fig:histogram} shows histograms of the distribution of reddening values for the different samples. For all tracers, the peaks of the distributions correspond to reddening values inferior to 0.1 mag. The slope of the distributions towards higher reddening values is however different among the different tracers. It is shallower for the dust emission based reddening than for the stellar ones. The ETG reddening does not show a prominent tail, but rather a wide peak where the distribution of values is confined within about 0.15 mag for the majoirty of the sources. 

\begin{figure}
\centering
\includegraphics[width=\columnwidth]{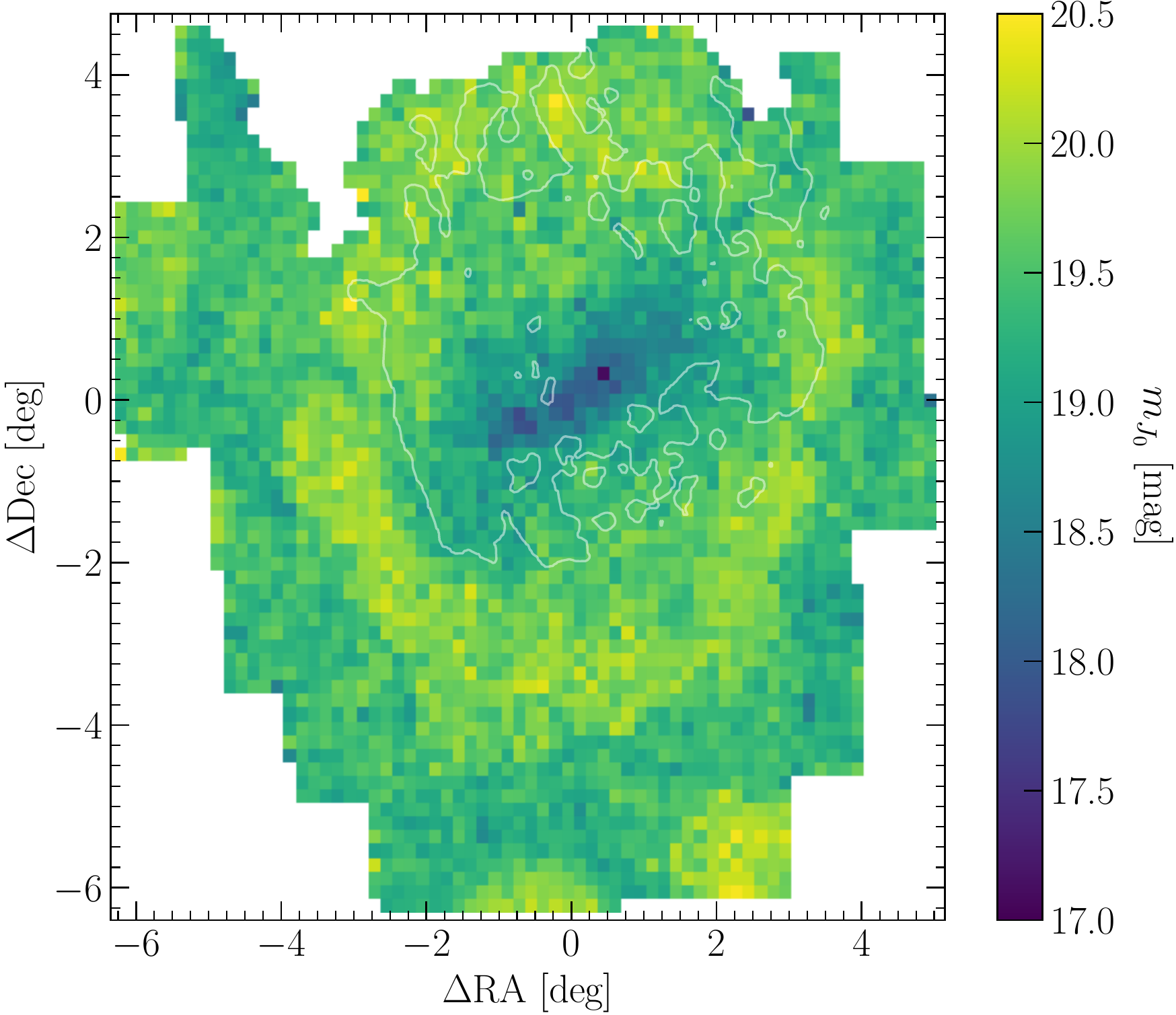} 
\caption[]{Map of the ETG extinction corrected J-band magnitude ($m_{J_{\mathrm{0}}}$) across the LMC at a resolution of 10$\times$10 arcmin$^2$.}
\label{fig:mag_map}
\end{figure}

Figure \ref{fig:mag_map} shows the magnitude of ETGs after applying the reddenign correction. This figure was created in the same way as that described for the reddening map, but instead of taking the median value of all best-fit reddening values in a given bin, we take the median of all extinction corrected J-band magnitudes. Note that the J-band magnitudes are calculated from the LAMBDAR fluxes and corrected for extinction using the relation shown in eqn.~1 of Paper~I. We use the J-band magnitude because it is available for almost all ETGs (essentially 100\%; 222736 out of 222752 ETGs), whereas for the optical bands we notice a decrease in the number of ETGs with measured fluxes that grows as one moves to bluer bandpasses, e.g. only 49\% and 88\% of ETGs have measured u- and g-band fluxes, respectively. In the r-, i- and z-bands, we have 97\%, 98\% and 99\% completeness, respectively. The variation of the magnitude of ETGs shows that they sample a different physical scale and therefore a different amount of dust towards the inner and outer regions of the LMC. This is due to the limitation by crowding in the centre and the shallow exposures in the outermost area. More line of sights are needed to describe the likely extinction in these areas compared to that in the intermediate area where the magnitude of ETGs is fairly homogeneous at about J=20.2 mag. 

\section{Summary}
\label{summary}

In this study, we have determined the total intrinsic reddening across $\simeq$\,90\,deg$^{2}$ of the LMC based on the analysis of SEDs of background galaxies. We followed a similar technique as developed in Paper I and extended to the SMC in Paper II. The main steps involved in our procedure and the conclusions are as follows.

\begin{itemize}
\item[(i)] We select $\sim$\,2.5 million background sources from the SMASH and VMC catalogues of the LMC by combining colour-magnitude and morphological criteria. We use \textsc{lambdar} to measure fluxes and construct SEDs from the optical ($ugriz$) to the near-IR ($YJK_{\rm{s}}$).
   \item[(ii)] We run the \textsc{lephare} $\chi^{2}$ code to fit the SEDs of the objects, which include 21,828 spectroscopically confirmed AGN, and select 252,752 galaxies with low levels of intrinsic reddening. The resulting reddening map reproduces the pervasive enhanced levels of reddening across the LMC bar region with enhancements  associated with star forming regions and overdensities traced by young stars.
    \item[(iii)] The calculated \textsc{lephare} photometric redshifts for 189 AGN are in very good agreement with the spectroscopically determined redshifts  ($\Delta z=z_{\rm{spec}}-z_{\rm{phot}}=-0.009$).
   \item[(iv)] We compare our reddening map with publicly available maps of the LMC. For the inner region, we find good agreement between our map and the distribution of dust emission. The comparison with stellar tracers, however, is more complicated owing to the variations amongst the reddening maps of the various stellar populations. For those showing a significant level of reddening associated with 30 Dor and molecular ridge we find discrepancies. Given the reduced number of galaxies in highly extinguished and crowded regions, it is possible our map is biased towards lower levels of reddening. In contrast, we find agreement with those maps demonstrating reddening in the bar region and in the outskirts of the galaxy. Given a factor of 2 difference, by sampling the full line-of-sight, and the uncertainties in the reddening determinations our map is consistent with maps derived from RC stars. Furthermore, the good agreement with a SFH-based reddening map indicates that a significant fraction of the dust is located beyond the bar.
   \item[(v)] We find that the regions lacking $u$-band data, mostly located in the outskirts of the galaxy, sample ETGs similar to those within the same $griz$ magnitude ranges resulting from deep exposures.
\end{itemize}

In this study, we have extended and improved our technique to measure the total intrinsic reddening of the LMC using background galaxies. The sample of 252,752 ETGs we used may seem small compared to the size of stellar samples adopted for the same purpose, it is however extremely large if one considers that the LMC lies in front of it.
Future studies based on large spectroscopic surveys (including surveys using the 4-metre multi-object spectroscopic telescope (4MOST); e.g. \citealp{Cioni19}) will provide spectra for hundreds of thousands of background galaxies which will allow us to not only robustly determine their redshifts but also to develop a complementary method to quantify the total intrinsic reddening in the foreground Magellanic Clouds. 

\section*{Acknowledgements}

This project has received funding from the European Research Council (ERC) under the European Union's Horizon
2020 research and innovation programme (grant agreement no. 682115).
This research was supported in part by the Australian Research Council Centre of Excellence for All Sky Astrophysics in
3 Dimensions (ASTRO3D), through project number CE170100013.
Y.C. acknowledges support from NSF grant AST 1655677. A.N. acknowledges support from the Narodowe Centrum Nauki (UMO-2020/38/E/ST9/00077).
%
%
%
%
The authors would like to thank the Cambridge Astronomy Survey Unit (CASU) and the Wide Field Astronomy
Unit (WFAU) in Edinburgh for providing the necessary data products under the support of the Science and Technology
Facilities Council (STFC) in the U.K.
The authors would also like to thank S. Rubele for his work on creating the deep, near-IR VMC PSF photometric catalogues.
This study was based on observations made with VISTA at the La Silla Paranal Observatory under programme ID 179.B-2003.
This project used data obtained with the
Dark Energy Camera (DECam), which was constructed by the
Dark Energy Survey (DES) collaboration.
Funding for the DES Projects has been provided by 
the U.S. Department of Energy, 
the U.S. National Science Foundation, 
the Ministry of Science and Education of Spain, 
the STFC, 
the Higher Education Funding Council for England, 
the National Center for Supercomputing Applications at the University of Illinois at Urbana-Champaign, 
the Kavli Institute of Cosmological Physics at the University of Chicago, 
the Center for Cosmology and Astro-Particle Physics at the Ohio State University, 
the Mitchell Institute for Fundamental Physics and Astronomy at Texas A\&M University, 
Financiadora de Estudos e Projetos, Funda{\c c}{\~a}o Carlos Chagas Filho de Amparo {\`a} Pesquisa do Estado do Rio de Janeiro, 
Conselho Nacional de Desenvolvimento Cient{\'i}fico e Tecnol{\'o}gico and the Minist{\'e}rio da Ci{\^e}ncia, Tecnologia e Inovac{\~a}o, 
the Deutsche Forschungsgemeinschaft, 
and the Collaborating Institutions in the Dark Energy Survey. 
The Collaborating Institutions are 
Argonne National Laboratory, 
the University of California at Santa Cruz, 
the University of Cambridge, 
Centro de Investigaciones En{\'e}rgeticas, Medioambientales y Tecnol{\'o}gicas-Madrid, 
the University of Chicago, 
University College London, 
the DES-Brazil Consortium, 
the University of Edinburgh, 
the Eidgen{\"o}ssische Technische Hoch\-schule (ETH) Z{\"u}rich, 
Fermi National Accelerator Laboratory, 
the University of Illinois at Urbana-Champaign, 
the Institut de Ci{\`e}ncies de l'Espai (IEEC/CSIC), 
the Institut de F{\'i}sica d'Altes Energies, 
Lawrence Berkeley National Laboratory, 
the Ludwig-Maximilians Universit{\"a}t M{\"u}nchen and the associated Excellence Cluster Universe, 
the University of Michigan, 
{the} National Optical Astronomy Observatory, 
the University of Nottingham, 
the Ohio State University, 
the University of Pennsylvania, 
the University of Portsmouth, 
SLAC National Accelerator Laboratory, 
Stanford University, 
the University of Sussex  
and Texas A\&M University.
Based on
observations at Cerro Tololo Inter-American Observatory,
National Optical Astronomy Observatory (NOAO Prop. ID:
2013A-0411 and 2013B-0440; PI: Nidever), which is operated
by the Association of Universities for Research in Astronomy
(AURA) under a cooperative agreement with the National
Science Foundation.
Finally, this project has made extensive use of the Tool for OPerations on Catalogues And
Tables (TOPCAT) software package \citep{Taylor05} as well as the following open-source
\texttt{Python} packages: Astropy \citep{Astropy18}, matplotlib \citep{Hunter07}, NumPy
\citep{Oliphant15}, pandas \citep{McKinney10}, and SciPy \citep{Jones01}.

\section*{Data Availability }

The data underlying this study are available in the study and in its online supplementary material.
The SMASH survey images used to create the galaxy SEDs are publicly available at the
NOIRLab Astro Data Lab (\url{https://datalab.noirlab.edu/}). The deep stack VMC images
of the LMC will be released in 2022, whereas the individual observations are publicly available
at the ESO archive (\url{http://archive.eso.org/cms.html}).
The galaxy SEDs will be shared on reasonable request to the corresponding author.

\bibliographystyle{mn3e}
\bibliography{references}

\label{lastpage}

\end{document}